\documentclass[transmag]{IEEEtran}

\ifCLASSOPTIONcompsoc
\usepackage[caption=false,font=normalsize,labelfont=sf,textfont=sf]{subfig}
\else
\usepackage[caption=false,font=footnotesize]{subfig}
\fi

\usepackage{xcolor}
\definecolor{highlight}{rgb}{0,0,0}
\definecolor{color}{rgb}{0,0,0}
\definecolor{color_I}{rgb}{0,0,0}
\definecolor{color_II}{rgb}{0,0,0}
\usepackage[hidelinks]{hyperref}
\usepackage{latexsym}
\usepackage{graphicx}
\usepackage{amsfonts,amssymb,amsmath}
\usepackage{multirow}
\usepackage{stfloats}
\usepackage{diagbox}
\usepackage{bm}
\usepackage{float}
\usepackage{cite}
\usepackage{algorithm}
\usepackage{algorithmic}
\usepackage{url}
\usepackage{epsfig}
\usepackage{graphicx}
\usepackage{makecell}
\usepackage{footmisc}
\usepackage{caption}
\usepackage[para]{threeparttable}
\newcommand{\tabincell}[2]{\begin{tabular}{@{}#1@{}}#2\end{tabular}} 
\usepackage{hyperref}
\def\BibTeX{{\rm B\kern-.05em{\sc i\kern-.025em b}\kern-.08em T\kern-.1667em\lower.7ex\hbox{E}\kern-.125emX}}
\begin{document}
	
	\title{PowerFDNet: Deep Learning-Based Stealthy False Data Injection Attack Detection for AC-model Transmission Systems}
	
	\author{Xuefei~Yin,
		Yanming~Zhu,
		Yi Xie,
		and~Jiankun~Hu$^{*}$,~\IEEEmembership{Senior Memeber,~IEEE}% <-
		
		\thanks{Xuefei Yin is with the School of Engineering and Information Technology, University of New South Wales, Canberra, ACT 2600, Australia (e-mail:  xuefei.yin@unsw.edu.au).}
		
		\thanks{Yanming Zhu is with the School of Computer Science and Engineering, University of New South Wales, Sydney, NSW 2052, Australia (e-mail:  yanming.zhu@unsw.edu.au).}
		
		\thanks{Yi Xie is with the School of Data and Computer Science (and the GuangDong Province Key Laboratory of Information Security Technology), Sun Yat-sen University, Guangzhou 510006, P.R. China (e-mail: xieyi5@mail.sysu.edu.cn).}
		
		\thanks{($\ast$ Corresponding author) Jiankun Hu is with the School of Engineering and Information Technology, University of New South Wales, Canberra, ACT 2600, Australia (e-mail:  j.hu@adfa.edu.au).} 
	}
	
	%The abstract should not exceed 250 words.
	\IEEEtitleabstractindextext{
		\begin{abstract}
			Smart grids are vulnerable to stealthy false data injection attacks (SFDIAs), as SFDIAs can bypass residual-based bad data detection mechanisms. 
			Methods based on deep learning technology have shown promising accuracy in the detection of SFDIAs.
			However, most existing methods rely on the temporal structure of a sequence of measurements but do not take account of the spatial structure between buses and transmission lines. 
			To address this issue, we propose a spatiotemporal deep network, PowerFDNet, for the SFDIA detection in AC-model power grids. The PowerFDNet consists of two sub-architectures: spatial architecture (SA) and temporal architecture (TA). The SA is aimed at extracting representations of bus/line measurements and modeling the spatial structure based on their representations. The TA is aimed at modeling the temporal structure of a sequence of measurements. 
			Therefore, the proposed PowerFDNet can effectively model the spatiotemporal structure of measurements. Case studies on the detection of SFDIAs on the benchmark smart grids show that the PowerFDNet achieved significant improvement compared with the state-of-the-art SFDIA detection methods. In addition, an IoT-oriented lightweight prototype of size 52 MB is implemented and tested for mobile devices, which demonstrates the potential applications on mobile devices.
			The trained model will be available at \textit{https://github.com/HubYZ/PowerFDNet}.
		\end{abstract}
		
	}
	
	\maketitle
	
	%Please use ``soft'' (e.g., \verb|\eqref{Eq}|) cross references instead
	%of ``hard'' references (e.g., \verb|(1)|). 
	%Please don't use the \verb|{eqnarray}| equation environment. Use
	%\verb|{align}| or \verb|{IEEEeqnarray}| instead. 
	
	\section{Introduction}\label{sec:introduction}
	\IEEEPARstart{I}{n} smart grids, a stealthy false data injection attack (SFDIA), which is aimed at maliciously manipulating measurements in a smart grid and may lead to serious consequences for the power system \cite{RN2089, RN1309, RN2085}, has recently become a focus on smart grid research \cite{RN2158, RN878, RN883, RN1312, RN2678}. 
	% \cite{RN2087,RN2086,RN2088,RN1310,RN2123}
	%RN877,RN2024,
	Contrary to other cyber-attacks (such as jamming and distributed denial-of-service), studies have proved that well-defined SFDIA measurements can bypass traditional bad measurement detection mechanisms \cite{RN877, RN878}, as the attacks obey power equations. 
	To address this issue, machine learning-based detection approaches have been explored and have obtained promising detection results \cite{RN2118, RN1100, RN2098, RN2092, RN2101}. 
	%RN2096,RN2099,RN2113,RN2112,RN2091
	Those experimental results demonstrate that deep learning can model the relationships between measurements and state variables in power grids to a certain extent. 
	This also promotes the further exploration of deep learning technology in this field \cite{RN2092}.
	
	Most of the existing machine learning-based methods are specifically designed for DC power systems \cite{RN2118, RN1100, RN2101}. 
	%Ozay \textit{et al.} \cite{RN2118} observed machine learning algorithms such as k-NN, SVM, Adaboost, and MKL in the detection of SFDIA for DC-model power grids. He \textit{et al.} \cite{RN1100} developed a deep-learning network for the detection of electricity theft in DC-model power grids. Wang \textit{et al.} \cite{RN2101} proposed a deep-learning-based method to locate SFDIAs in DC-model power grids. 
	However, as the DC-model power system is a simplified AC-model system, the estimated states of the DC-model cannot truly represent the actual states of the AC-model power system. Therefore, those methods are not well-suited for the SFDIA detection in real-world AC-model power systems.
	In recent years, researchers have explored the application of deep learning in AC-model power systems \cite{RN2104, RN2106, RN2110, RN2111, RN2090, RN2097}. 
	Kundu \textit{et al.} \cite{RN2110} proposed an SFDIA detection method based on an auto-encoder to attempt to capture the relations between system states and measurements. 
	Zhang \textit{et al.} \cite{RN2097} extended this auto-encoder scheme by embedding a generative adversarial network. 
	%Both the two methods treat the measurements as one-dimensional input feature. That would result in losing the spatial structure of the measurements. 
	Contrary to these two methods, Yu \textit{et al.} \cite{RN2090} proposed a detection method, which instead uses state values obtained from measurements as the input feature.
	
	Although those AC-based methods propose effective models to represent the relationship between measurements and the states of power grids, they focus on learning the temporal structure of a sequence of measurements but do not take account of the spatial structure of the measurements. 
	The spatial structure refers to the relationship of the measurements (/state variables) between buses and transmission lines; %\cite{RN878}
	the temporal structure refers to the relationship of the measurements (/state variables) collected at a continuous time \cite{RN877}. 
	The method proposed by Kundu \textit{et al.} \cite{RN2110} mainly modeled the temporal structure of a sequence of measurements using an auto-encoder. 
	%A sequence of measurements captured at a continuous time is used to train the recurrent neural network (RNN) auto-encoder for modeling temporal structure information. 
	As a disadvantage, this method did not consider the spatial structure information between buses and transmission lines in the power grid, because the measurements are mixed into a one-dimensional input. 
	Similarly, a deep network approach proposed in \cite{RN2097} also takes the measurements as one-dimensional inputs and leads to the loss of the spatial structure information between transmission lines and buses. 
	Contrary to the aforementioned two methods, Yu \textit{et al.} \cite{RN2090} proposed a deep neural network (DNN) based detector using gated recurrent units (GRUs) to learn the temporal structure information of a sequence of measurements. 
	One of the major differences between the aforementioned two methods is that this method trains the detection network by taking as input the state values calculated from the measurements. 
	In summary, the aforementioned methods can model the temporal structure information by applying GRUs, auto-encoders, and recurrent neural networks. However, the spatial structure information between transmission lines and buses is not considered in these methods, resulting in limited detection accuracy.
	
	To address this issue, this paper proposes a spatiotemporal deep learning network (named PowerFDNet) for SFDIA detection in AC power systems.
	The proposed PowerFDNet contains two key sub-architectures: spatial architecture (SA) and temporal architecture (TA). 
	The SA aims to extract representations from bus/line measurements and model the spatial structure of the measurements by learning their representations. 
	The TA aims to learn the temporal structure of time-series measurements and make the decision. 
	To naturally model the spatiotemporal structure information, measurements with and without SFDIAs are utilized as the training data. 
	The comprehensive experiments evaluated on the benchmark power grids demonstrate that the proposed PowerFDNet achieves significant improvement in detection accuracy ($F_1$, recall, and precision) in comparison with the state-of-the-art methods. 
	
	The main contributions of this paper are summarized as follows:
	\begin{itemize}
		\item[1)] Compared to most existing approaches that mainly model temporal structure information for SFDIA detection, we propose a new network to learn the global spatiotemporal structure information of measurements.  
		\item[2)] Compared to most existing approaches that take measurements as one-dimensional input, we propose well-designed residual-based sub-networks to learn multidimensional representations for buses and lines separately. 
		% 	the features of each bus/line and 
		Specifically, we propose using well-designed convolutional layers and residual connections to model the multidimensional representations. 
		As an advantage, this facilitates the subsequent spatiotemporal structure learning. 
		\item[3)] To capture the temporal structure of the representations of a sequence of measurements, we design a booster-refiner feature encoder based on long short-term memory (LSTM) architecture. 
		The booster-refiner encoder first models the bus/line measurement data relationship with rich features and then refines the high-dimensional feature. 
		\item[4)] We generate and release a comprehensive SFDIA dataset for facilitating research works in this area. The SFDIA dataset is generated for the power grids in the SimBench dataset \cite{RN2073}, which contains a wide variety of power grids with high, medium, and low voltage, as well the corresponding load and generator profiles in 15-minute resolution for a whole year. 
		%	To the best of our knowledge, most, if not all, existing works are evaluated on the simplified IEEE test power system models. 
		Our released dataset can provide a more realistic evaluation setting. Therefore, with this SFDIA dataset, researchers in this area can focus on the design and analysis of SFDIA detection.
		\item[5)] An IoT-oriented lightweight prototype of size around 52 MB, with an optimized mobile model of size around 8.5 MB, is implemented and tested for mobile devices, which demonstrates the potential applications on mobile devices.
	\end{itemize}
	
	The rest of this paper is organized as follows: Section \ref{sec:relatedworks} reviews related works on deep learning-based SFDIA detection; Background knowledge about state estimation, bad data detection, and the SFDIA are presented in Section \ref{sec:background}; Section \ref{sec:proposedmethod} describes the proposed PowerFDNet in detail; the experiments and results are organized in Section \ref{sec:experimentalresults}; and at the end, we concluded the paper in Section \ref{sec:conclusion}.
	
	\section{Related Works} \label{sec:relatedworks}
	Bad data detection is one of the essential functions of estate estimation to detect measurement errors. Those measurements' errors may occur due to various reasons, such as the finite accuracy of meters, the telecommunication medium, and meters' failure \cite{RN2017, RN1221, RN1311}. Liu \textit{et al.} \cite{RN2141} validated that well-defined error data, as known as stealthy false data injection attack (SFDIA), can bypass the residual-based bad measurement detection in DC power grids; and in 2012, Hug \textit{et al.} \cite{RN878} established this type of attack to AC power grids. 
	%	To detect the SFDIA, various methods are developed \cite{RN1081}. 
	%	However, most of the works aim to defend against SFDIA by protecting or encrypting measurements \cite{RN1081}. 
	
	Some methods managed to detect SFDIAs by statistical methods \cite{RN2142, RN2143}, sparse optimization \cite{RN2145}, graph theory \cite{RN2148}, Kalman filter \cite{RN2144}, time-series simulation \cite{RN2149}, state forecasting \cite{RN2146, RN2147}, and machine learning \cite{RN1100, RN2150, RN2151, RN2118, RN2152, RN2101}. For example, Ozay \textit{et al.} \cite{RN2118} investigated SVM-based algorithms to classify measurements as being either secure or attacked. 
	The experimental results demonstrate that machine learning algorithms perform better on SFDIA detection than detection algorithms that employ state vector estimation. 
	He \textit{et al.} \cite{RN1100} proposed a real-time SFDIA detection method based on restricted Boltzmann machines \cite{RN1258}. 
	The advantage is that historical measurements are used to capture features for SFDIA detection. 
	However, this method did not take account of the spatial structure information between transmission lines and buses. 
	Wang \textit{et al.} \cite{RN2101} proposed a CNN-based method to detect SFDIA attacks. The advantage is that this method attempts to identify the attack locations. However, this method failed to consider the temporal structure information.
	Besides, the aforementioned methods are mainly developed to detect SFDIAs in DC-model power systems. 
	
	In recent years, with the development of deep learning techniques, researchers have explored the application of deep learning to AC model power systems \cite{RN2104, RN2106, RN2097, RN2110, RN2111, RN2090}. 
	Kundu \textit{et al.} \cite{RN2110} presented a detection approach based on an auto-encoder by modeling the relations between system states and measurements. 
	The advantage is that historical measurements are used to detect SFDIAs. 
	The disadvantage is that it did not take into account the relationship between line measurements and bus measurements. 
	Zhang \textit{et al.} \cite{RN2097} extended this auto-encoder scheme by embedding a generative adversarial network. 
	One of the common points between the two methods is that the measurements are taken as the input feature. 
	Different from that, Yu \textit{et al.} \cite{RN2090} proposed an SFDIA detection method, which instead utilizes state variables (i.e., bus voltage angles and magnitudes) as the input feature. 
	However, that incurs two potential risks. 
	One is that the original spatiotemporal structure of measurements may be lost, as the model is learned from the estimated state variables instead of the measurements. 
	Another one is that state variables estimated from false measurements may be incorrect to the power grids at the current time. 
	Recently, Yin \textit{et al.} \cite{RN2678} proposed a sub-grid-oriented
	microservice-based supervising network through privacy-preserving collaborative learning to detect SFDIAs. 
	However, this method mainly considered the local spatiotemporal relationship of measurement in sub-grids in the privacy-preserving setting.
	The work will focus on modeling the global spatiotemporal structure in a non-privacy-preserving setting.

	% To address this issue, this study proposes a spatiotemporal deep learning approach for the SFDIA detection in AC-model power grids. Specifically, the spatial structure between transmission lines and buses is firstly modeled by a spatial architecture, and then a temporal architecture is proposed to capture the temporal structure information of a sequence of measurements.
	
	\section{Background} \label{sec:background}
	In this section, we firstly provide the brief background knowledge related to the residual-based bad measurement data detection and the SFDIA. Then, we introduce an approach to generate the SFDIAs against AC-model power grids. Some key notations in this paper are listed in Table \ref{tab:notations}.
	\begin{table}[htbp]
		%	\vspace{-0.2in}
		\setlength\tabcolsep{4pt}
		\renewcommand{\arraystretch}{1.5}
		\caption{Key notations}
		%	\vspace{0.1in}
		\label{tab:notations}
		\centering
		%	\vspace{-0.08in}
		\begin{tabular}{c|l} 
			\hline
			\hline
			\textbf{Symbol} &\textbf{Comments}
			\\ \hline 
			SFDIAs& stealthy false data injection attacks
			\\ \hline 
			SA & spatial architecture
			\\ \hline
			TA & temporal architecture
			\\ \hline
			$\bm{z}$ & meter measurements
			\\ \hline
			$\bm{x}$ & state variables
			\\ \hline
			$\bm{\varepsilon}$ & measurement errors
			\\ \hline
			$P_{ik}$ & active power flow from bus i to bus k
			\\ \hline
			$Q_{ik}$ & reactive power flow from bus i to bus k
			\\ \hline
			$P_i$ & active power injection at bus i
			\\ \hline
			$Q_i$ & reactive power injection at bus i
			\\ \hline
			$\bm{r}$ & measurement residual
			\\ \hline
			$\bm{H(x)}$ & power functions of the state variables
			\\ \hline
			$m_b$ & the number of monitored buses
			\\ \hline
			$c_b$ & the maximum number of measurements at each monitored bus
			\\ \hline
			$m_l$ & the number of monitored lines
			\\ \hline
			$c_l$ & the maximum number of measurements at each monitored lines
			\\ \hline
			$T$ & a time window
			\\ \hline
			$t_k$ & a time step
			\\ \hline
			$\bm{z}_{t_k}^{b}$ & bus measurements collected at time $t_k$
			\\ \hline
			$\bm{z}_{t_k}^{l}$ & line measurements collected at time $t_k$
			\\ \hline
			$\bm{Z}_{t_k}$ & a sequence of measurements collected in the time window $T$
			
			\\ 	\hline 	\hline
		\end{tabular}
	\end{table}
	
	\subsection{AC State Estimation} \label{sec:stateEstimation}
	The major objective of the state estimation is to determine optimal power states (i.e., bus voltage and angle) based on a set of redundant measurements for the power system \cite{RN2017}. 
	The measurements are usually comprised of bus measurements and line measurements. 
	The bus measurements typically consist of active/reactive power injection (i.e., bus load and generation) and bus voltage magnitude. 
	The line measurements typically consist of active/reactive power flows measured at two sides of the transmission lines and line current flow magnitudes. 
	In an AC-model power grid, the nonlinear formulation between the state variable $\bm{x}$ and the measurement $\bm{z}$ can be expressed by \cite{RN2017}: 
	\begin{equation} \label{eq:hx}
		\bm{z = H(x) + \varepsilon},
	\end{equation}
	where $\bm{z} \in \mathbb{R}^{m}$ denotes the measurements, $\bm{\varepsilon} \in \mathbb{R}^{m}$ denotes the errors of the measurements, $\bm{x} = [\theta_{2},\theta_{3},\cdots,\theta_{n},V_1,V_2, \cdots, V_n]^T \in \mathbb{R}^{2n-1}$ denotes bus voltage angles and magnitudes for a $n$-bus power grid, and $\bm{H(x)}$ is the nonlinear vector function of the state variables. 
	$h_i(\bm{x}) \in \bm{H(x)}$ is the formula of the measurement $z_i$ related to the state variable $\bm{x}$. 
	It is assumed that the error $\varepsilon_{i} \in \bm{\varepsilon}$ is independent and draws from a Gaussian distribution $\mathcal{N}(0,\sigma_{i}^{2})$. 
	
	The nonlinear functions of each measurement and power states are presented as follows. The active and reactive power flows from bus $i$ to bus $k$, $P_{ik}$ and $Q_{ik}$, can be expressed by \cite{RN2017}
	\begin{equation}\label{eq:powerfow}
		\begin{cases}
			P_{ik} = V^2_i(g_{ik}+g_{si}) - V_{i}V_{k}(b_{ik}\sin\theta_{ik} + g_{ik} \cos\theta_{ik}), \\
			Q_{ik}= -V^2_i(b_{ik} - b_{si}) - V_{i}V_{k}(g_{ik}\sin\theta_{ik}-b_{ik}\cos\theta_{ik}), \\
		\end{cases}
	\end{equation}
	and, line current flow magnitude from bus $i$ to bus $k$, $I_{ik}$, can be expressed by \cite{RN2017}
	\begin{equation}\label{eq:currentflow}
		I_{ik} = \sqrt{P_{ik}^2 + Q_{ik}^2} / Vi.
	\end{equation}
	The active and reactive power injections at bus $i$, $P_{i}$ and $Q_{i}$, can be expressed by
	\begin{equation}\label{eq:powerinjection}
		\begin{cases}
			P_{i} = V_i \sum_{k \in \Omega_i} V_k(B_{ik} \sin \theta_{ik} + G_{ik} \cos\theta_{ik}), \\
			Q_{i} = V_i \sum_{k \in \Omega_i} V_k(G_{ik} \sin\theta_{ik} - B_{ik} \cos \theta_{ik}),
		\end{cases}
	\end{equation}
	where $\theta_i$ and $V_i$ represent the state variables for bus $i$, $\theta_k$ and $V_k$ represent the state variables for bus $k$, $\theta_{ik} = \theta_i - \theta_k$, $g_{ik}+jb_{ik}$ is the admittance of the line connecting buses $i$ and $k$, and $g_{si}+jb_{si}$ is the admittance of the shunt line at bus $i$. 
	$G_{ik} +jB_{ik}$ is the $ik$th element of its complex bus admittance matrix. 
	$\Omega_i$ denotes the set of adjacent buses that are directly connected to bus $i$.
	
	The state estimation is to find the optimal solution $\bm{\hat{x}}$ based on the measurements through minimizing the following weighted least squares problem: 
	\begin{equation} \label{eq:Jx}
		\bm{\mathcal{L}(x) = (z-H(x))^{T} \varLambda^{-1} (z-H(x))},
	\end{equation}
	where $\bm{\varLambda} = diag[\sigma^{2}_{1}, \sigma^{2}_{2}, \cdots, \sigma^{2}_{m}]$ denotes the weight matrix whose element represents the variance of the measurements at the corresponding electricity meter. 
	%	For conventional meters, $\sigma_{i} = (0.02a + 0.0052b)/3$, where $a$ denotes the measured value and $b$ denotes the full scale value \cite{RN2069}. 
	The minimization of the objective function $\bm{\mathcal{L}(x)}$ can be solved by iterative approaches (e.g., the Newton-Raphson algorithm), which can be expressed by
	\begin{equation} \label{eq:minJx}
		\hat{\bm{x}} = \arg\min_{\bm{x}} \bm{\mathcal{L}(x)},
	\end{equation}
	where $\bm{\hat{x}}$ denotes the optimal power system state estimated from the measurements.
	
	\subsection{Residual-based Bad Measurement Detection} \label{sec:badDataDetection}
	Errors in the measurements may be introduced due to distinct reasons such as meter malfunction, signal transmission interference, and cyberattacks. 
	The bad data detection is aimed at determining whether or not the measurements contain significant errors. 
	The values of state variables estimated from normal meter measurements should be close to the true state values, while the state values estimated from bad measurements may be significantly different from the true values. Thus, one popular method is to calculate the residual between the estimated measurements $\bm{H(\hat{x})}$ and the observed measurements $\bm{z}$, which can be formulated by 
	\begin{equation} \label{eq:r}
		\bm{r = z - H(\hat{x})}.
	\end{equation}
	If the residual is greater than a threshold, the measurements are treated as bad measurements. 
	\textit{Chi-square} test \cite{RN2017} is commonly used to decide the threshold.  
	Specifically, $\dfrac{r_{i}}{\sigma_{i}}$ follows the standard normal distribution, where $r_{i} \in \bm{r}$.
	\begin{equation} \label{eq:chi-square}
		\varUpsilon = \sum_{i=1}^{m}(\dfrac{r_{i}}{\sigma_{i}})^2.
	\end{equation}
	Then, $\varUpsilon$ follows a $m-(2n-1)$ degrees of freedom Chi-squared distribution. 
	Based on the theory of the Chi-squared test, the threshold $\tau$ is determined by a hypothesis test with a significance level $\alpha$ \cite{RN2017}. Therefore, with the probability $\alpha$, the presence of bad measurements is inferred if 
	\begin{equation}
		\varUpsilon \geq \tau^2.
	\end{equation}
	
	\subsection{Stealthy False Data Injection Attack} \label{sec:falseDataInjectionAttack}
	Stealthy false data injection attacks are aimed at circumventing the bad data detection mechanism by deliberately manipulating some measurements. The stealthy attack is designed based on the bad data detection mechanism \cite{RN878}. Let $\bm{z_{bad}}$ denote the measurements maliciously modified by an SFDIA, which can be expressed by
	\begin{equation}\label{eq:zbad}
		\bm{z_{bad} = \bm{z} + a},
	\end{equation}
	and $\bm{\hat{x}_{a}}$ denote the corresponding power states estimated from $\bm{z_{bad}}$, which can be expressed by
	\begin{equation}
		\bm{\hat{x}_a = \hat{x}+c}.
	\end{equation}
	Hence, we have Eq. \eqref{eq:attack} \cite{RN878},
	\begin{equation} \label{eq:attack}
		\begin{split}
			\bm{\| z_{bad} - H(\hat{x}_{a}) \|} &= \bm{ \| z+a - H(\hat{x}+c) \|} \\
			&=\left\|
			\begin{bmatrix}
				\bm{z_1}\\
				\bm{z_2 + a_2}
			\end{bmatrix}
			- 
			\begin{bmatrix}
				\bm{H_1(\hat{x}_1)}\\
				\bm{H_2(\hat{x}_1,\hat{x}_2+c_2)}
			\end{bmatrix}
			\right\|
		\end{split}
	\end{equation}
	where variables with subscript `$\bm{1}$' indicate that they will not be modified by the attack, while variables with subscript `$\bm{2}$' need to be maliciously modified.
	The vectors $\bm{a_2}$ and $\bm{c_2}$ are the changes on the measurements and state variables, respectively.
	Hence, if $\bm{a_2}$ is obtained by Eq. \eqref{eq:attack_vector},  
	\begin{equation}\label{eq:attack_vector}
		\bm{a_2 = H_2(\hat{x}_1, \hat{x}_2+c_2) - H_2(\hat{x}_1,\hat{x}_2)},
	\end{equation} 
	Eq. \eqref{eq:attack} can then be expressed as follows \cite{RN878}
	\begin{equation}		
		\begin{split}
			\bm{\| z_{bad} - H(\hat{x}_{a}) \|} 
			&=\left\|
			\begin{bmatrix}
				\bm{z_1}\\
				\bm{z_2}
			\end{bmatrix}
			- 
			\begin{bmatrix}
				\bm{H_1(\hat{x}_1)}\\
				\bm{H_2(\hat{x}_1,\hat{x}_2)}
			\end{bmatrix}
			\right\|
			\\
			&=\bm{\| z-H(\hat{x})\|}
			\\
			&=\bm{r}.
		\end{split}
	\end{equation}
	Therefore, the malicious attack measurement obtained by Eq. \eqref{eq:attack_vector} can bypass the detection mechanism.

	\section{Proposed PowerFDNet} \label{sec:proposedmethod}
	In this section, we present details of the proposed PowerFDNet to detect SFDIAs in AC-model power grids. 
	%The SFDIA detection is formulated as a binary classification problem. 
	The PowerFDNet is aimed at classifying the measurements to determine whether measurements are maliciously modified, which consists of two key sub-architectures: a spatial architecture (SA) and a temporal architecture (TA). 
	The SA aims to extract representations from bus/line measurements and model the spatial structure of measurements by learning their representations (Section \ref{sec:SA}). 
	The TA aims to model the temporal structure of the measurements by learning the intermediate feature obtained by the SA and making a final prediction (Section \ref{sec:TA}). 
	%Fig. \ref{fig:SA-V} provides an overview of the architecture of the proposed PowerFDNet. 
	%\begin{figure}[htbp]
	%	%	\vspace{-0.3in}
	%	\centering
	%	\includegraphics[width=0.95\linewidth]{figures/SA-V.eps}
	%	\caption{The architecture of the proposed PowerFDNet for SFDIA detection in AC-model power systems. $T$ stands for the time window. $f_{c-n-a}$ denotes a convolution followed by normalization and activation, where the subscript $c$ denotes the convolution, $n$ denotes the normalization, and $a$ denotes the activation. $f_{l-a}$ indicates a linear perceptron followed by activation, where $l$ indicates the linear perceptron. $ks$ in the dashed rectangles denotes the kernel size; $O$ denotes the size of output features; $I$ denotes the size of input features. The operator $P$ means reshaping a tensor; $S$ means squeezing a tensor by removing the dimensions of size 1; $C$ means concatenating a tensor; and $L$ means selecting the data in the last row.}
	%	\label{fig:SA-V}
	%\end{figure}
	
	\subsection{Measurement Data} \label{sec:measureData}
	In this subsection, we provide the details about the organization of the measurements. Common measurement variables and notations such as in \cite{RN2090, RN2017, RN2678} are adopted.
	Typically, the measurements for power systems include line measurements and bus measurements. 
	The line measurements commonly contain active and reactive power flow data measured at the two sides of transmission lines and line current flow magnitudes, which are summarized as follows: 
	\begin{itemize}
		\item $P_I$, active power flow measurement at the `\textit{in}' side,
		\item $P_O$, active power flow measurement at the `\textit{out}' side,
		\item $Q_I$, reactive power flow measurement at the `\textit{in}' side,
		\item $Q_O$, reactive power flow measurement at the `\textit{out}' side,
		\item $I_I$, current flow magnitude measurement at the `\textit{in}' side,
		\item $I_O$, current flow magnitude measurement at the `\textit{out}' side.
	\end{itemize}
	The typical bus measurements are summarized as follows: 
	\begin{itemize}
		\item $P$, bus active power injection measurement,
		\item $Q$, bus reactive power injection measurement,
		\item $V$, bus voltage magnitude measurement.
	\end{itemize}
	
	Let $\bm{z}_{t_k}^{b} \in \mathbb{R}^{m_{b}\times 1\times c_b}$ denote the bus measurements collected at time $t_k$, where $m_b$ denotes the number of monitored buses and $c_b$ denotes the maximum number of measurements at each monitored bus. 
	Similarly, let $\bm{z}^{l}_{t_k} \in \mathbb{R}^{m_l \times  1\times c_l}$ denote the line measurements collected at time $t_k$, where $m_l$ denotes the number of monitored lines and $c_l$ denotes the maximum number of measurements at each monitored line. Measurements with less than $c_b$/$c_l$ data on monitored buses/lines are padded with 0.
	Hence, the measurements of a power grid collected at time $t_k$ can be expressed by $\bm{z}_{t_k}$, which is composed of $\bm{z}_{t_k}^{b}$ and $\bm{z}^{l}_{t_k}$, formulated by 
	$\bm{z}_{t_k} = \{ \bm{z}_{t_k}^{b}, \bm{z}^{l}_{t_k}\}.$
	Let
	$\bm{Z}_{t_k} = \{\bm{z}_{t_{k-T}}, \bm{z}_{t_{k-(T-1)}}, \cdots, \bm{z}_{t_{k-1}}, \bm{z}_{t_k}\} $ denote a sequence of measurements,
	where $T$ is a constant value. 
	For convenience, $\bm{Z}_{t_k}$ is also equivalently expressed by  
	$\bm{Z}_{t_k} = \{\bm{Z}^{b}_{t_k}, \bm{Z}^{l}_{t_k}\},$
	where $$\bm{Z}^{b}_{t_k} = \{\bm{z}^b_{t_{k-T}}, \bm{z}^b_{t_{k-(T-1)}}, \cdots, \bm{z}^b_{t_k}\}\in \mathbb{R}^{T\times m_b \times 1 \times c_b}$$ is the time-series bus measurements and 
	$$\bm{Z}^{l}_{t_k} = \{\bm{z}^l_{t_{k-T}}, \bm{z}^l_{t_{k-(T-1)}}, \cdots, \bm{z}^l_{t_k}\}\in \mathbb{R}^{T\times m_l \times 1 \times c_l}$$ is the time-series line measurements.
	We define the label for $\bm{Z}_{t_k}$ as follows:
	\begin{equation} 
		\label{eq:y}
		y_{t_k} = 
		\begin{cases}
			0 , & \mbox{if } \bm{z}_{t_k} \mbox{ is normal measurement data }\\
			& ~~~\mbox{that has not been attacked;}\\
			1,  & \mbox{if } \bm{z}_{t_k} \mbox{ is attacked by an SFDIA and } \\ &~~~\mbox{can bypass the residual-based detection.}
		\end{cases}
	\end{equation}
	In this paper, the commercial power system analysis software PowerFactory 2017 SP4 \footnote{https://www.digsilent.de/en/powerfactory.html} was used to conduct the residual-based bad measurement detection.
	
	\subsection{Spatial Architecture (SA)} \label{sec:SA}
	
	\begin{figure*}[htbp] 
		\centering
		\includegraphics[width=0.8\linewidth]{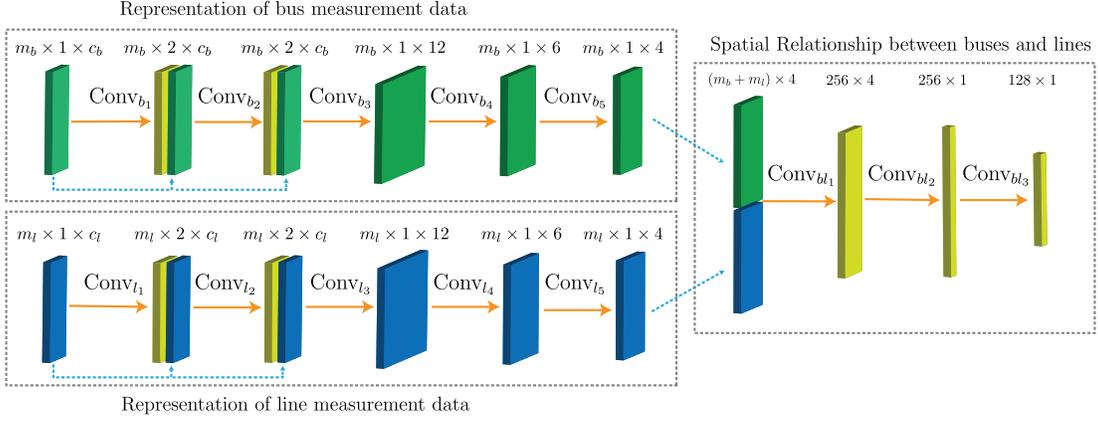}
		\caption{The SA network architecture. The configuration of the SA is shown in Table \ref{tab:bus_representation}, Table \ref{tab:line_representation}, and Table \ref{tab:spatial_relationship}.}
		\label{fig:SA}
	\end{figure*}
	
	%\begin{figure}[htbp] 
	%	\centering
	%	\includegraphics[width=0.99\linewidth]{figures/Bus_denoise_R.eps}
	%	\caption{A 12-dimensional feature is transformed into a 4-dimensional feature by the perceptron layer $f_{l-a}$.}
	%	%	\vspace{-0.15in}
	%	\label{fig:bus_denoise_R}
	%\end{figure}
	%
	%\begin{figure}[htbp] 
	%	\centering
	%	\includegraphics[width=0.99\linewidth]{figures/Line_denoise_R.eps}
	%	\caption{A 12-dimensional feature is transformed into a 4-dimensional feature by the perceptron layer $f_{l-a}$.}
	%	%	\vspace{-0.15in}
	%	\label{fig:line_denoise_R}
	%\end{figure}
	
	The SA is aimed at modeling the spatial structure between buses and lines by learning the representations of their measurements. 
	The architecture of SA is shown in Fig. \ref{fig:SA}. 
	It is composed of three sub-networks: one for bus measurement representation, one for line measurement representation, and the other one for spatial structure modeling.
	Different from the methods \cite{RN1100, RN2101, RN2110, RN2097, RN2090} that squeeze line and bus measurements into a one-dimensional input, we propose extracting the representations for lines and buses separately. 
	As an advantage, the SA can effectively learn the hidden features from bus and line measurements and model their spatial structures. 
	Compared with the method in \cite{RN2678}, we utilize residual blocks to extract the bus/line representations and deploy deep layers to make them suitable for modeling the global spatial structure of the large-scale power grids.
	
	%Both of them contains operations $f_{c-n-a}$ and $f_{l-a}$. 
	%$f_{c-n-a}$ denotes a convolution followed by a normalization and an activation, where the subscript $c$ denotes the convolution, $n$ denotes the normalization, and $a$ denotes the activation. $f_{l-a}$ indicates a linear perceptron followed by an activation, where $l$ indicates the linear perceptron.
	
	\subsubsection{Sub-network for Representation of Bus Measurement Data}\label{sec:bus_representation}
	This sub-network is aimed at extracting representations of bus measurements. The architecture is shown in Fig. \ref{fig:SA}.
	The parameters are summarized in Table \ref{tab:bus_representation}. To learn 
	To learn the residual from the measurements \cite{RN377}, the intermediate features are concatenated with the original measurement in the second and third layers. The fourth and fifth layers are aimed at extracting the representations of bus measurements from the intermediate feature space. 
	\begin{table}[htbp]
		%	\vspace{-0.2in}
		\setlength\tabcolsep{2pt}
		\renewcommand{\arraystretch}{1.5}
		\caption{Configuration of the sub-network for the representation of bus measurements}
		%	\vspace{0.1in}
		\label{tab:bus_representation}
		\centering
		%	\vspace{-0.08in}
		\begin{threeparttable}
			\resizebox{\linewidth}{!}
			{
				\begin{tabular}{c|c|c|c|c|c|c} 
					\hline
					\hline
					\textbf{Layer} &\textbf{\tabincell{c}{Kernel \\ Size}} &\textbf{\tabincell{c}{Out \\Channels}} & \textbf{Groups} & \textbf{Stride} & \textbf{Padding} & \textbf{BN/AF/Reshape} 
					\\ \hline 
					%			Layer 				Kernel size				Out channel		Groups			Stride			Padding     BN/AF
					Conv$_{b_1}$&$1\times c_b$	& $m_b$				& $m_b$ 	& 1 			& [0, 1]	 & Yes/ELU/No
					\\ \hline
					Conv$_{b_2}$&$2\times c_b$	& $m_b$				&$m_b$		& 1				& [0, 1]	 &Yes/ELU/No
					\\ \hline
					Conv$_{b_3}$&$2\times c_b$	& 12$m_b$	      &$m_b$	& 1				& [0, 0]	 &Yes/ELU/Yes
					\\ \hline
					Conv$_{b_4}$&$1\times3$		& $m_b$	  	  		&$m_b$		& 2	  			& [0, 1]	&Yes/ELU/No
					\\ \hline
					Conv$_{b_5}$&$1\times3$		 & $m_b$	  	  	&$m_b$	  	& 1				& [0, 0]	 &Yes/ELU/Yes
					\\ 	\hline 	\hline
				\end{tabular}
			}
			\begin{tablenotes}
				\setlength\tabcolsep{6pt}
				\renewcommand{\arraystretch}{1}
				\footnotesize
				\begin{tabular}{ll}
					\textbf{BN}: Batch normalization & 
					\textbf{AF}: Activation function
				\end{tabular}
			\end{tablenotes}
			
		\end{threeparttable}
		%		\vspace{-0.2in}
	\end{table}
	
	Specifically, for bus measurement $\bm{Z}^{b}_{t_k}$, the first layer Conv$_{b_1}$ utilizes $c_b$ convolution filters of size $1\times c_b$ to extract a residual feature for each bus, formulated by 
	\begin{equation}
		\bm{b}^1_{t_k,o_j} =  \mbox{Conv}_{b_1}(\bm{z}^b_{t_k}) = \phi(\varphi(\bm{W}_{o_j} \ast \bm{z}^b_{t_k} + a_{o_j})),
	\end{equation}
	where $a_{o_j}$ is an additive bias for each output channel, $\bm{W}_{o_j} \in \mathbb{R}^{1\times c_b}$, $\bm{z}_{t_k}^{b} \in \mathbb{R}^{m_{b}\times 1 \times c_b}$, $\bm{b}^1_{t_k,o_j} \in \mathbb{R}^{m_b\times 1}$, and $\ast$ denotes the convolutional operator. $\varphi$ denotes the batch normalization, and $\phi$ denotes the exponential linear unit (ELU) \cite{RN2153}. Then, the residual feature and the bus measurements are concatenated into a two-channel data, which are fed into the next layer. The last layer $\mbox{Conv}_{b_5}$ utilizes four $1\times6$ convolution filters to extract the representation for each bus. 
	Therefore, a representation of four feature values is obtained for each bus.
	For the input bus measurements $\bm{Z}^{b}_{t_k}$, its representation is denoted by $\bm{B}_{t_k} \in \mathbb{R}^{T\times m_b \times 1 \times 4} $.

	\subsubsection{Sub-network for Representation of Line Measurement Data}\label{sec:line_representation}
	
	This sub-network is proposed to extract the representations for the line measurements. 
	The network architecture is shown in Fig. \ref{fig:SA}.
	The parameters are summarized in Table \ref{tab:line_representation}.
	In the second and third layers, the previous results are concatenated with the original measurement to learn the residual from the line measurements. 
	The fourth and fifth layers are designed to learn the line representations from the intermediate feature space. 
	
	\begin{table}[htbp]
		%	\vspace{-0.2in}
		\setlength\tabcolsep{2pt}
		\renewcommand{\arraystretch}{1.5}
		\caption{Configuration of the sub-network for the representation of line measurements}
		%	\vspace{0.1in}
		\label{tab:line_representation}
		\centering
		%	\vspace{-0.08in}
		\begin{threeparttable}
			\resizebox{\linewidth}{!}
			{
				\begin{tabular}{c|c|c|c|c|c|c} 
					\hline
					\hline
					\textbf{Layer} &\textbf{\tabincell{c}{Kernel \\ Size}} &\textbf{\tabincell{c}{Out \\Channels}} & \textbf{Groups} & \textbf{Stride} & \textbf{Padding} & \textbf{BN/AF/Reshape} 
					\\ \hline 
					%				Layer 				Kernel size				Out channel		Groups			Stride			Padding     BN/AF
					Conv$_{l_1}$&$1\times c_l$  & $m_l$				& $m_l$ 	& 1 			& [0, 1]	 & Yes/ELU/No
					\\ \hline
					Conv$_{l_2}$&$2\times c_l$ & $m_l$				&$m_l$		& 1				& [0, 1]	 &Yes/ELU/No
					\\ \hline
					Conv$_{l_3}$&$2\times c_l$ & 12$m_l$	      &$m_l$	& 1				& [0, 0]	 &Yes/ELU/Yes
					\\ \hline
					Conv$_{l_4}$&$1\times 3$& $m_l$	  	  		&$m_l$		& 2	  			& [0, 1]	&Yes/ELU/No
					\\ \hline
					Conv$_{l_5}$&$1\times 3$& $m_l$	  	  	&$m_l$	  	& 1				& [0, 0]	 &Yes/ELU/Yes
					\\ 	\hline 	\hline
				\end{tabular}
			}
			\begin{tablenotes}
				\setlength\tabcolsep{6pt}
				\renewcommand{\arraystretch}{1}
				\footnotesize
				\begin{tabular}{ll}
					\textbf{BN}: Batch normalization & 
					\textbf{AF}: Activation function
				\end{tabular}
			\end{tablenotes}
			
		\end{threeparttable}
		%		\vspace{-0.2in}
	\end{table}
	%	
	%\begin{table}[htbp]
	%	%	\vspace{-0.2in}
	%	\setlength\tabcolsep{4pt}
	%	\renewcommand{\arraystretch}{1.5}
	%	\caption{Configuration of parameters in the sub-network for the representation of line measurements}
	%	%	\vspace{0.1in}
	%	\label{tab:line_representation}
	%	\centering
	%	%	\vspace{-0.08in}
	%	\begin{tabular}{c|c|c|c|c} 
	%		\hline
	%		\hline
	%		\textbf{Convolution} &\textbf{\tabincell{c}{Kernel \\ Size}} & \textbf{\tabincell{c}{Kernel \\Numbers}} & \textbf{\tabincell{c}{Batch \\Normalization}} & \textbf{\tabincell{c}{Activation\\Function}}
	%		\\ \hline 
	%		Conv$_{l_1}$&$1\times c_l$& $c_l$ &Yes & ELU
	%		\\ \hline
	%		Conv$_{l_2}$&$1\times c_l$& $c_l$ &Yes & ELU
	%		\\ \hline
	%		Conv$_{l_3}$&$1\times c_l$& 12 &Yes & ELU
	%		\\ \hline
	%		Conv$_{l_4}$&$1\times12$& 6&Yes & ELU
	%		\\ \hline
	%		Conv$_{l_5}$&$1\times6$& 4&Yes & ELU
	%		\\ 	\hline 	\hline
	%	\end{tabular}
	%\end{table}
	
	Similar to the sub-network for the bus representation, a representation of four feature values is learned for each line.
	Therefore, for the input line measurements $\bm{Z}^{l}_{t_k}$, its representation is denoted by $\bm{L}_{t_k} \in \mathbb{R}^{T\times m_l \times 1 \times 4} $.
	
	\subsubsection{Sub-network for Modeling Spatial Structure}
	This sub-network is proposed to learn the spatial structure of measurements from the measurement representations, as shown in Fig. \ref{fig:SA}.
	The parameters are summarized in Table \ref{tab:spatial_relationship}.
	To learn the spatial structure between each line/bus and the remaining ones, we propose utilizing larger filters with the size of $(m_b+m_l)\times 1$. 
	Therefore, output features learned from the bus/line representations can reflect such relationships. 
	\begin{table}[htbp]
		%	\vspace{-0.2in}
		\setlength\tabcolsep{4pt}
		\renewcommand{\arraystretch}{1.5}
		\caption{Configuration of the sub-network for modeling the spatial structure}
		%	\vspace{0.1in}
		\label{tab:spatial_relationship}
		\centering
		%	\vspace{-0.08in}
		\begin{threeparttable}
			\resizebox{\linewidth}{!}
			{
				\begin{tabular}{c|c|c|c|c|c|c} 
					\hline
					\hline
					\textbf{Layer} &\textbf{\tabincell{c}{Kernel \\ Size}} &\textbf{\tabincell{c}{Out \\Channels}} & \textbf{Groups} & \textbf{Stride} & \textbf{Padding} & \textbf{BN/AF/Reshape} 
					\\ \hline 
					%Layer 				Kernel size							Out channel		Groups			Stride			Padding     BN/AF
					Conv$_{l_1}$&$(m_b+m_l)\times 1$  & 256					& 1					& 1 			& [0, 0]	 	& Yes/ELU/No
					\\ \hline
					Conv$_{l_2}$&$1\times 4$ 				  & 256					&256			 & 1			  & [0, 0]	       &Yes/ELU/Yes
					\\ 	\hline 	
					Conv$_{l_3}$&$256\times 1$  			& 128				 & 1					& 1 			& [0, 0]	 	& Yes/ELU/Yes
					\\ \hline					
					\hline
				\end{tabular}
			}
			\begin{tablenotes}
				\setlength\tabcolsep{6pt}
				\renewcommand{\arraystretch}{1}
				\footnotesize
				\begin{tabular}{ll}
					\textbf{BN}: Batch normalization & 
					\textbf{AF}: Activation function
				\end{tabular}
			\end{tablenotes}
			
		\end{threeparttable}
		%		\vspace{-0.2in}
	\end{table}
	
	%	\begin{table}[htbp]
	%		%	\vspace{-0.2in}
	%		\setlength\tabcolsep{2pt}
	%		\renewcommand{\arraystretch}{1.5}
	%		\caption{Configuration of parameters in the sub-network for modeling the spatial structure}
	%		%	\vspace{0.1in}
	%		\label{tab:spatial_relationship}
	%		\centering
	%		%	\vspace{-0.08in}
	%		\begin{tabular}{c|c|c|c|c} 
	%			\hline
	%			\hline
	%			\textbf{Convolution} & \textbf{\tabincell{c}{Kernel \\ Size}} & \textbf{\tabincell{c}{Kernel \\Numbers}} & \textbf{\tabincell{c}{Batch \\Normalization}} & \textbf{\tabincell{c}{Activation\\Function}}
	%			\\ \hline 
	%			Conv$_{bl_1}$&$1\times 4$& $ 512$ & Yes & ELU
	%			\\ \hline
	%			Conv$_{bl_2}$&$256 \times 1$& $128$ & Yes & ELU 
	%			\\ \hline
	%			Conv$_{bl_3}$&$128 \times 1$& $64$  & Yes & ELU
	%			\\ 	\hline 	\hline
	%		\end{tabular}
	%	\end{table}
	
	Specifically, the bus and line representations are firstly reshaped and concatenated to fully represent the input measurements, expressed by $$\bm{H}_{t_k} = \{\bm{B}_{t_k}, \bm{L}_{t_k}\} \in \mathbb{R}^{T\times 1 \times (m_b+m_l) \times 4}.$$
	Then, three layers are designed to learn the spatial structure of line/bus measurements, which is expressed by 
	\begin{equation}
		\bm{S}_{t_k} = \mbox{Conv}_{bl_3}(\mbox{Conv}_{bl_2}(\mbox{Conv}_{bl_1}(\bm{H}_{t_k}))),
	\end{equation}
	where $$\bm{S}_{t_k} = \{\bm{s}_{t_{i-T}}, \bm{s}_{t_{i-(T-1)}}, \cdots,\bm{s}_{t_{i-1}},  \bm{s}_{t_k}\} \in \mathbb{R}^{T\times 128}.$$
	
	\subsection{Temporal Architecture (TA)} \label{sec:TA}
	\begin{figure*}[htbp] 
		\centering
		\includegraphics[width=0.7\linewidth]{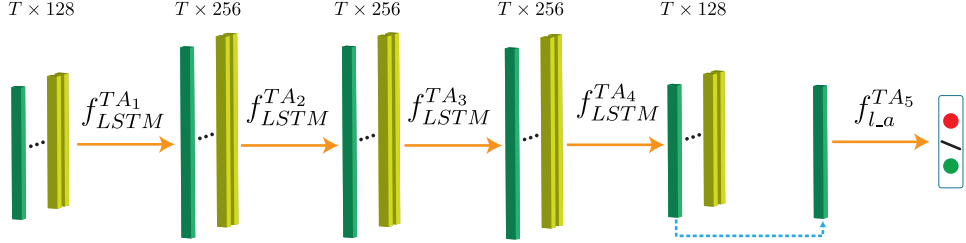}
		\caption{The TA network architecture. The configuration of the TA is shown in Table \ref{tab:temporal_relationship}.}
		%	\vspace{-0.15in}
		\label{fig:TA}
	\end{figure*}
	
	\begin{table}[htbp]
		%	\vspace{-0.2in}
		\setlength\tabcolsep{15pt}
		\renewcommand{\arraystretch}{1.5}
		\caption{Configuration of parameters in the TA for modeling the temporal structure}
		%	\vspace{0.1in}
		\label{tab:temporal_relationship}
		\centering
		%	\vspace{-0.08in}
		\begin{tabular}{c|c|c} 
			\hline
			\hline
			\textbf{Layer} &\textbf{Input Size} & \textbf{Hidden Size} 
			\\ \hline 
			$f^{TA_1}_{LSTM}$&128& 256
			\\ \hline
			$f^{TA_2}_{LSTM}$&256& 256
			\\ \hline
			$f^{TA_3}_{LSTM}$&256 & 256
			\\ \hline
			$f^{TA_4}_{LSTM}$&256& 128
			\\ 	\hline 	\hline
		\end{tabular}
	\end{table}
	
\begin{figure}[htbp]
		\centering
		\includegraphics[width=0.85\linewidth]{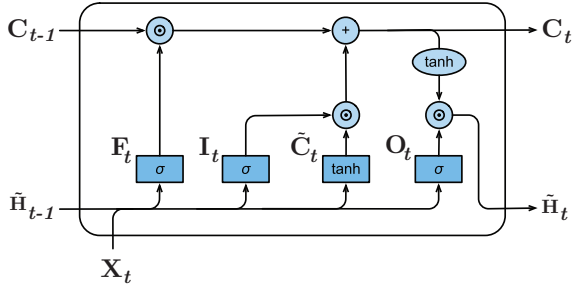}
		\caption{The architecture of the LSTM \cite{zhang2020dive}. $\bm{C}_{t-1}$ is the cell state, $\bm{\tilde{H}}_{t-1}$ is the hidden state, $\bm{F}_{t}$ is the forget gate, $\bm{I}_{t}$ is the input gate, $\bm{\tilde{C}}_{t}$ is the candidate memory, and $\bm{O}_{t}$ is the output gate. $\sigma$ denotes the sigmoid function and $\odot$ denotes the Hadamard product.
		}
		\label{fig:LSTM}
		%	\vspace{-0.15in}
	\end{figure}

	The TA is used to model the temporal structure of a sequence of measurements in a time window by learning their intermediate features obtained by the SA. 
	The architecture of TA is shown in Fig. \ref{fig:TA}. 
	It is composed of four LSTM layers, one fully connected layer, and a sigmoid layer. 
	The output represents the probability that the measurement $\bm{z}_{t_k}$ at time step $t_k$ is an SFDIA. 
	To effectively model the temporal structure information, we incorporate the LSTM architecture shown in Fig. \ref{fig:LSTM} into the proposed PowerFDNet. 
	It has two major advantages: one is to represent temporal structure information of time-series measurements \cite{RN2128}, and another is to avoid gradient vanishing and exploding \cite{RN1214}. This architecture forms a booster-refiner encoder that can use rich features to model the large-scale spatial representations of bus and line measurement variables and then refines them with a more salient and condensed feature vector representation. 
	
	Recall that for the input measurements $\bm{Z}_{t_k}$
	%$$\bm{Z}_{t_k} = \{\bm{z}_{t_{i-T}}, \bm{z}_{t_{i-(T-1)}}, \cdots, \bm{z}_{t_{i-1}}, \bm{z}_{t_k}\},$$
	the spatial structure information 
	$\bm{S}_{t_k}$
	is learned by the SA. 
	For convenience, the input data for time step $t$ is denoted by $\bm{X}_t \in \bm{S}_{t_k}$, $d=128$, and $h = 256$.
	Specifically, the calculation flow in the $f^{TA_1}_{LSTM}$ is expressed as follows: 
	$$\bm{F}_t = \sigma(\bm{X}_t \bm{W}_{xf} + \bm{\tilde{H}}_{t-1} \bm{W}_{hf} + \bm{b}_f), $$
	where $\bm{W}_{xf} \in \mathbb{R}^{d\times h}, \bm{W}_{hf} \in \mathbb{R}^{h\times h}, 
	\bm{b}_f \in \mathbb{R}^{1 \times h}$, and $\sigma$ denotes the sigmoid function;
	$$\bm{I}_t = \sigma(\bm{X}_t \bm{W}_{xi} + \bm{\tilde{H}}_{t-1} \bm{W}_{hi} + \bm{b}_i),
	$$
	where $\bm{W}_{xi} \in \mathbb{R}^{d\times h}, \bm{W}_{hi} \in \mathbb{R}^{h\times h}, 
	\bm{b}_i \in \mathbb{R}^{1 \times h};$
	$$\tilde{\bm{C}}_t = \text{tanh}(\bm{X}_t \bm{W}_{xc} + \bm{\tilde{H}}_{t-1} \bm{W}_{hc} + \bm{b}_c),$$
	where $ \bm{W}_{xc} \in \mathbb{R}^{d\times h}, \bm{W}_{hc} \in \mathbb{R}^{h\times h}, 
	\bm{b}_c \in \mathbb{R}^{1 \times h};$
	$$\bm{O}_t = \sigma(\bm{X}_t \bm{W}_{xo} + \bm{\tilde{H}}_{t-1} \bm{W}_{ho} + \bm{b}_o),
	$$ where $\bm{W}_{xo} \in \mathbb{R}^{d\times h}, 
	\bm{W}_{ho} \in \mathbb{R}^{h\times h}, 
	\bm{b}_o \in \mathbb{R}^{1 \times h};$
	$$\bm{C}_t = \bm{F}_t \odot \bm{C}_{t-1} + \bm{I}_t \odot \tilde{\bm{C}}_t,$$
	where $\bm{C}_{t-1} \in \mathbb{R}^{1 \times h}$ and $\odot$ denotes the Hadamard product. 
	Then, the output of the $f^{TA_1}_{LSTM}$ is expressed by
	\begin{equation}
		\bm{\tilde{H}}_t = \bm{O}_t \odot \tanh(\bm{C}_t).
	\end{equation}
	After the four LSTM layers, the feature map $\bm{X}_{out} \in \mathbb{R}^{T \times d}$ is obtained. 
	To detect the SFDIA, the feature data for the current time step is detached, represented by $\bm{x}_p \in \mathbb{R}^{1\times d}$. 
	This feature is then processed by layer $f^{TA5}_{l\_a}$ with a sigmoid activation, formulated by 
	\begin{equation} \label{eq:yp}
		y_p = \sigma(\bm{x}_p\bm{W}+a),
	\end{equation}
	where $\bm{W} \in \mathbb{R}^{d\times1}$ and $a \in \mathbb{R}$.
	
	\subsection{Loss Function}
	The binary cross entropy error is utilized to train the proposed PowerFDNet, which is expressed by
	\begin{equation}
		f_{loss} = -\sum_{i=1}^{N}(y_{i}\log y_{p_i} + (1-y_{i})\log (1-y_{p_i})),
	\end{equation}
	where $N$ is the mini-batch size, $y_{p_i}$ is the prediction result obtained by Eq. \eqref{eq:yp} for the measurements $\bm{z}_{t_k}$, and $y_{i}$ is the corresponding ground truth label (in Eq. \eqref{eq:y}). In the training stage, the optimization algorithm of Adam \cite{RN2155} is used to update the network weights, with an initial learning rate of $1\times10^{-4}$. The learning rate is dynamically adjusted in the training stage by the ReduceLROnPlateau scheduler.
	The popular deep learning framework Pytorch-1.9.0\footnote{https://pytorch.org/docs/1.9.0/} was used to construct the PowerFDNet for the model training and testing. 
	The trained network model of the proposed PowerFDNet will be available online at \textit{https://github.com/FrankYinXF/PowerFDNet}.
	
	\section{Case Studies} \label{sec:experimentalresults}
	
	\subsection{Dataset}
	In the experiments, two benchmark power systems from the public SimBench dataset \cite{RN2073} were used to assess 
	the SFDIA detection.
	% (instead of IEEE test power systems) 
	There are three main reasons.
	First, these power grids contain detailed data, especially time-series demand profiles for an entire year that are generated every 15 minutes (e.g., 35,136 demand profiles). Therefore, it can convenient to use these power grids to simulate a power grid with dynamical power load and generation.
	%Different from many methods \cite{RN2097, RN2090, RN2110, RN2101} that utilize IEEE test power systems without time series data, it is more reasonable and realistic to utilize the SimBench dataset for SFDIA detection. That is mainly because those methods must synthesize profiles for load and generation individually. 
	Second, measurements for buses and lines have been defined and tested in these SimBench power grids with high voltage and extra-high voltage \cite{RN2073}. Therefore, this dataset can be conveniently used to evaluate the SFDIA detection.
	Finally, the SimBench power grids originated from the German power systems. To some extent, it provides realistic power grids for evaluating the SFDIA detection.
	
	The two benchmark power grids, `\textit{1-HV-mixed--0-no\_sw}' and `\textit{1-EHV-mixed--0-no\_sw}', were utilized to evaluate the performance of SFDIA detection. 
	The `\textit{1-HV-mixed--0-no\_sw}' is a high voltage level grid with 110 KV transmission lines, denoted by Grid-HV, which is monitored by 355 measurements, with 35,136 profiles for dynamical power load and generation. 
	More details are shown in Table \ref{tab:Grid-HV}.
	\begin{table}[htbp]
		%	\vspace{-0.2in}
		\setlength\tabcolsep{4pt}
		\renewcommand{\arraystretch}{1.5}
		\caption{Details for the Grid-HV}
		%	\vspace{0.1in}
		\label{tab:Grid-HV}
		\centering
		%	\vspace{-0.08in}
		\begin{tabular}{c|c|l} 
			\hline
			\hline
			\textbf{Component} &\textbf{Quantity} & \textbf{Explanation}
			\\ \hline 
			Bus&64& all the buses are in service.
			\\ \hline%\Xhline{0.6pt}
			Load &58&
			\\ \hline%\Xhline{0.6pt}
			Lines & 95& all the lines are in service.
			\\ 	\hline
			Transformer & 6 &
			\\ 	\hline
			External grid & 3 &
			\\ 	\hline
			Bus measurements & 192 & measurement type: $P$, $Q$, and $V$
			\\ 	\hline
			Line measurements & 163 & measurement type: $P$, $Q$, and $I$
			\\ 	\hline
			\hline
		\end{tabular}
		%	\vspace{-0.1in}
	\end{table}
	The `\textit{1-EHV-mixed--0-no\_sw}' is an extra-high voltage level grid with 220-380 KV transmission lines, denoted by Grid-EHV, which is monitored by 3,952 measurements, with 35,136 profiles for dynamical power load and generation.
	More details are shown in Table \ref{tab:Grid-EHV}.
	\begin{table}[htbp]
		%	\vspace{-0.2in}
		\setlength\tabcolsep{4pt}
		\renewcommand{\arraystretch}{1.5}
		\caption{Details for the Grid-EHV}
		%	\vspace{0.1in}
		\label{tab:Grid-EHV}
		\centering
		%	\vspace{-0.08in}
		\begin{tabular}{c|c|l} 
			\hline
			\hline
			\textbf{Component} &\textbf{Quantity} & \textbf{Explanation}
			\\ \hline 
			Bus&571& all the buses are in service.
			\\ \hline%\Xhline{0.6pt}
			Load &390&
			\\ \hline%\Xhline{0.6pt}			
			Lines & 849& all the lines are in service.
			\\ 	\hline
			Transformer & 209 &
			\\ 	\hline
			External grid & 7 &
			\\ 	\hline
			Bus measurements & 1,698 & measurement type: $P$, $Q$, and $V$
			\\ 	\hline
			Line measurements & 2,254 & measurement type: $P$, $Q$, and $I$
			\\ 	\hline
			\hline
		\end{tabular}
		%	\vspace{-0.1in}
	\end{table}
	
	The open-source software Pandapower\footnote{https://www.pandapower.org/} and SimBench\footnote{https://simbench.readthedocs.io/en/stable/about/installation.html} and the commercial software PowerFactory 2017 SP4\footnote{https://www.digsilent.de/en/powerfactory.html} were used in the SFDIA date generation stage for the power flow calculation and the bad data detection.
	
	\subsubsection{Time-series Measurements Generated on the Grid-HV and Grid-EHV} \label{sec:measurement}
	The genuine values of these measurements in the two power grids were obtained by calculating the power flow using the commercial software PowerFactory 2017 SP4. 
	Hence, there are 35,136 normal measurement samples for each power grid. 
	{\color{color}Details of measurement data are presented in Section \ref{sec:measureData}.}
	For Grid-HV, each measurement sample collected at a time step contains 192 {\color{color}measurement} values for buses and 163 {\color{color}measurement} values for {\color{color} transmission lines}. 
	For Grid-EHV, each measurement sample measured at a time step contains 1,698 {\color{color}measurement} values for buses and 2,254 {\color{color}measurement} values for {\color{color} transmission lines}.
	The measurement noises are assumed to follow Gaussian distributions and are configured to be less than 1\% for voltage magnitude and less than 2\% for active/reactive power injection and power flow \cite{RN2017}.
	
	\subsubsection{SFDIA Measurement Generation} \label{sec:SFDIAgeneration}
\begin{figure}[htbp]
		\centering
		\includegraphics[width=\linewidth]{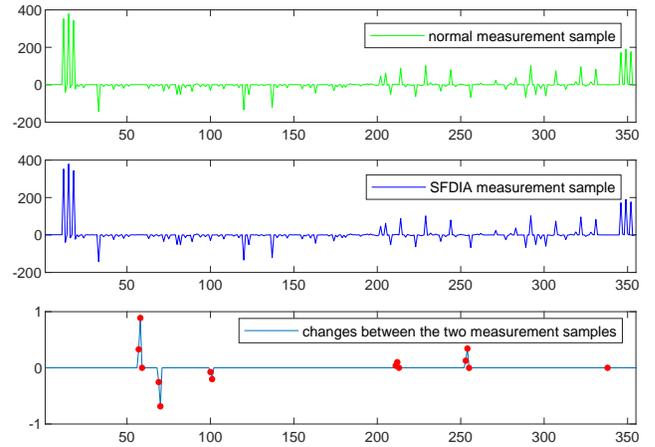}
		\caption{The x-axis is the measurement index, and the y-axis is the measurement value. The top figure shows a normal measurement sample ($\bm{z}$ in Eq. \eqref{eq:zbad}), the middle figure shows an SFDIA sample ($\bm{z_{bad}}$ in Eq. \eqref{eq:zbad}) obtained by attacking the normal sample, and the bottom figure shows the change ($\bm{a}$ in Eq. \eqref{eq:zbad}) between the two samples.}
		\label{fig:normal_SFDIA_samples}
		%	\vspace{-0.15in}
	\end{figure}

	The generation of the SFDIA measurement is based on the method proposed in \cite{RN878, RN2678}. The attacks were launched on a target bus by maliciously modifying either its voltage angle ($Va$) or voltage magnitude ($Vm$).
	%The changes of $Va$ are limited within the range of $\pm5^\circ$ and the changes of $Vm$ are within the range of $\pm0.02 p.u.$ \cite{RN878}. 
	To comprehensively evaluate the performance of the SFDIA detection, three types of SFDIAs are designed and summarized as follows:
	\begin{itemize}
		\item Type-A that the rate of the active power injection change on the target bus is in the range of $(50\%, 100\%]$,
		\item Type-B that the rate of the active power injection change on the target bus is in the range of $(25\%, 50\%]$, and
		\item Type-C that the rate of the active power injection change on the target bus is in the range of $(5\%, 25\%]$.
	\end{itemize}
	\begin{figure*}[b]
		\centering
		\subfloat[][]
		{
			\centering
			\includegraphics[width=0.3\linewidth]{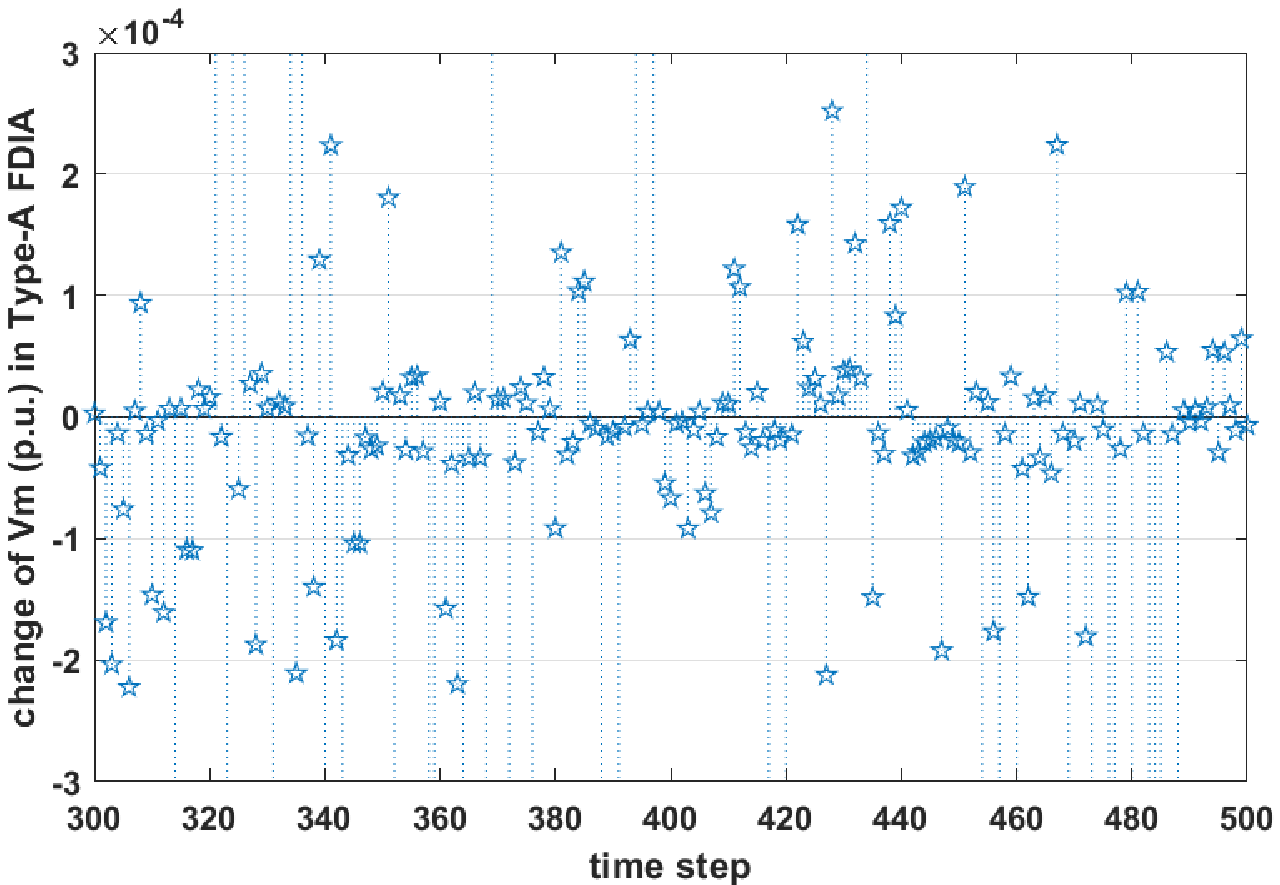}
			\label{fig:type_I_vm_change}
		}
		\hfil
		\subfloat[][]
		{
			\centering
			\includegraphics[width=0.3\linewidth]{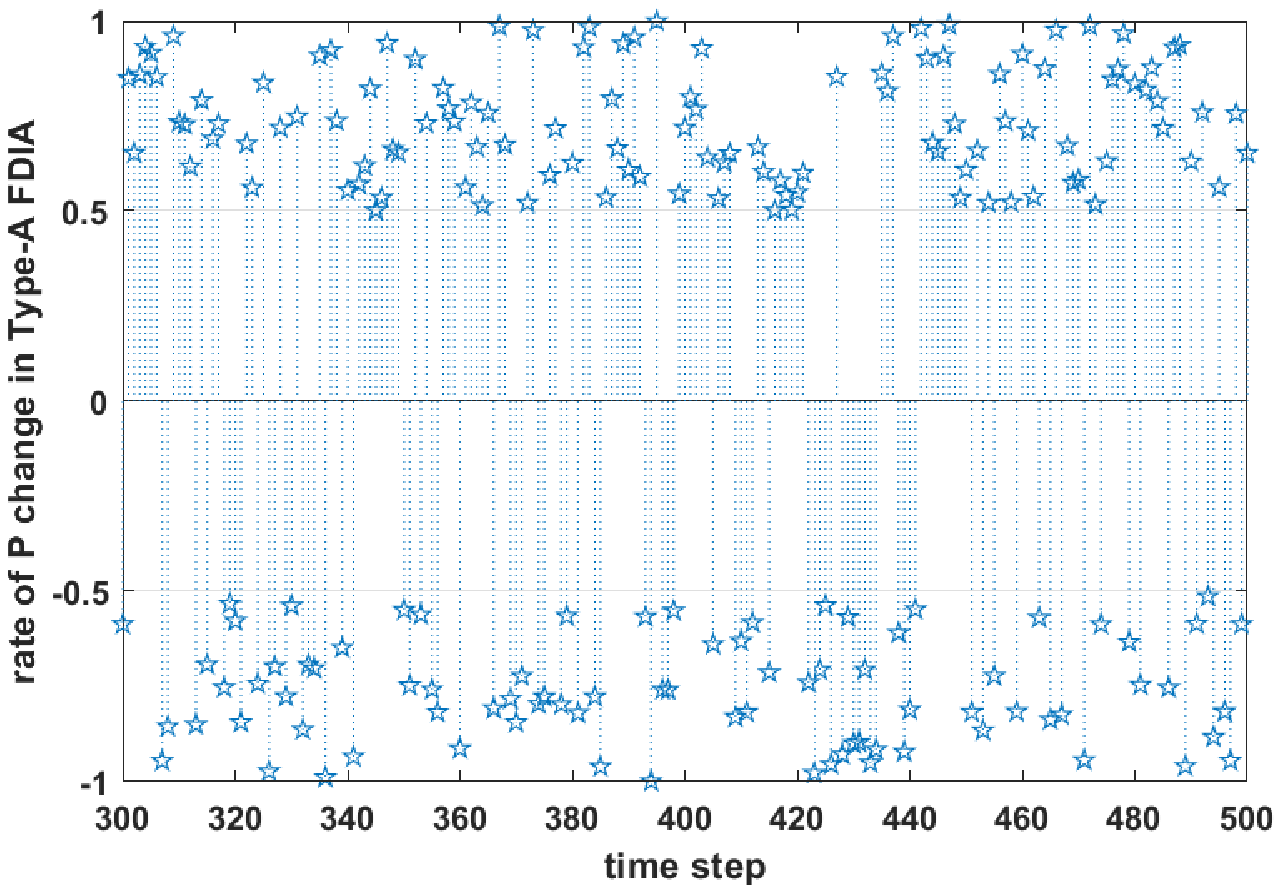}
			\label{fig:type_I_p_rate}
		}
		\hfil
		\subfloat[][]
		{
			\centering
			\includegraphics[width=0.3\linewidth]{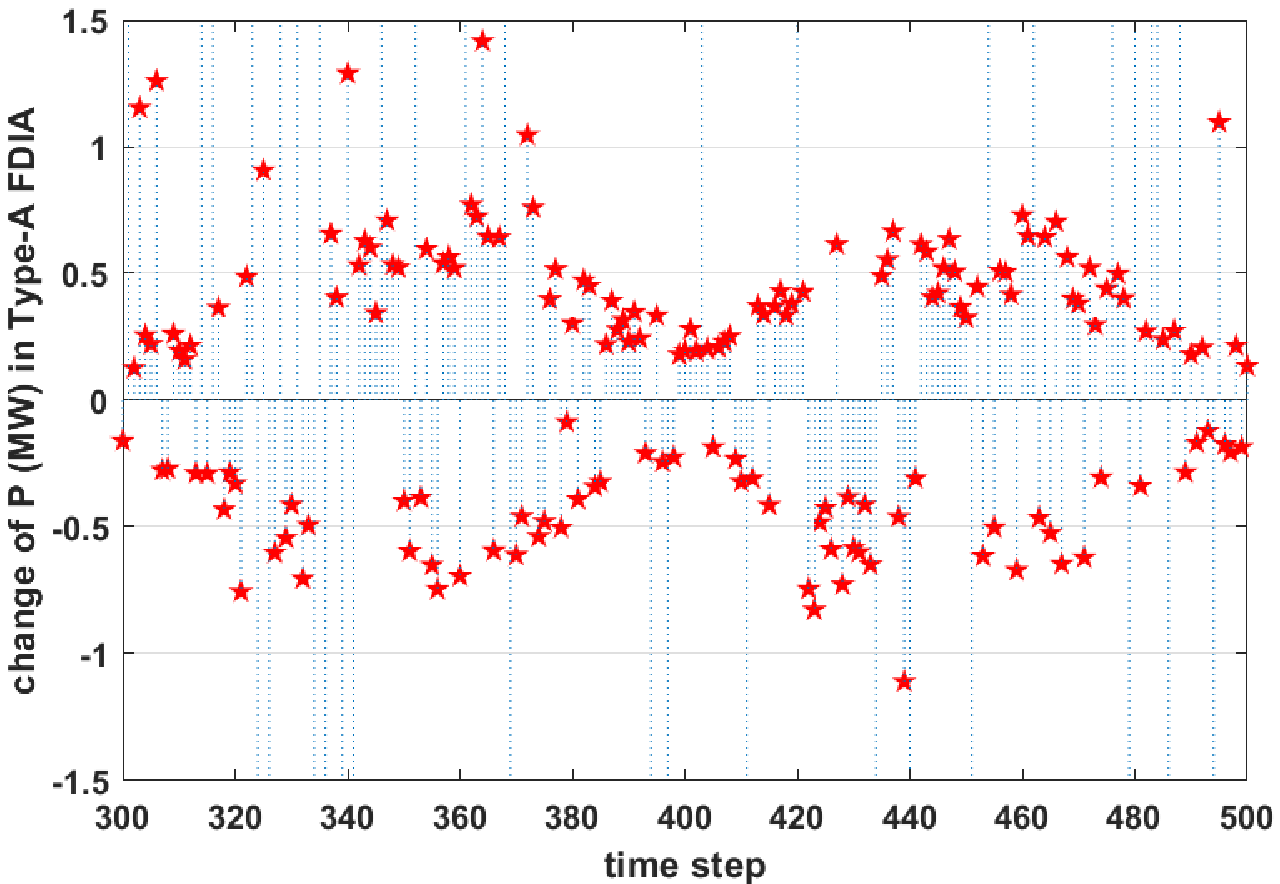}
			\label{fig:type_I_p_change}
		}
		\caption[]{Statistical information about Type-A SFDIA measurements at target buses in Grid-HV: \subref{fig:type_I_vm_change} the change of voltage magnitude, \subref{fig:type_I_p_rate} the distribution of the rate of activate power change, and \subref{fig:type_I_p_change} the change of activate power.}
		\label{fig:type_I}
	\end{figure*}
	Therefore, Type-A SFDIA will lead to a large change in power injection, Type-B SFDIA will lead to a medium change, and Type-C SFDIA will lead to a relatively small change in power injection.  
	At each time step, six buses with injection are randomly selected as the target buses to launch these three types of SFDIAs. 
	Therefore, there are totally 35,136 $\times$ 6 = 210,816 attacked measurement samples, summaries as follows: 
	\begin{itemize}
		\item 35,136 Type-A SFDIA attack measurements by attacking the bus magnitude $Vm$,
		\item 35,136 Type-A SFDIA attack measurements by attacking the bus angle $Va$,
		\item 35,136 Type-B SFDIA attack measurements by attacking the bus magnitude $Vm$,
		\item 35,136 Type-B SFDIA attack measurements by attacking the bus angle $Va$,
		\item 35,136 Type-C SFDIA attack measurements by attacking the bus magnitude $Vm$, and 
		\item 35,136 Type-C SFDIA attack measurements by attacking the bus angle $Va$.
	\end{itemize}
	All of these SFDIA measurements have bypassed the bad measurement detection function of PowerFactory 2017 SP4. 
	These three types of SFDIA datasets generated for Grid-HV and Grid-EHV in this experiment will be publicly available online at \textit{https://github.com/FrankYinXF/PowerFDNet}.
	Fig. \ref{fig:type_I} shows the statistical information from time step 300 to 500 about Type-A SFDIA attack measurements on Grid-HV in terms of the change of $Vm$, the rate of $P$ change, and the change of $P$.
	Fig. \ref{fig:normal_SFDIA_samples} shows a normal measurement sample (corresponding to $\bm{z}$ in Eq. \eqref{eq:zbad}), an SFDIA sample obtained (corresponding to $\bm{z_{bad}}$ in Eq. \eqref{eq:zbad}) by attacking the normal sample, and the change (corresponding to $\bm{a}$ in Eq. \eqref{eq:zbad}) between the two samples.

	\subsubsection{Training and Testing Dataset} \label{sec:trainingAndTesting}
	As introduced in Section \ref{sec:SFDIAgeneration}, each power grid generates 35,136 normal measurement samples and 210,816 SFDIA samples with three types of attacks. The normal measurements are labeled as $0$, and the SFDIA measurements are labeled as $1$, as expressed in Eq. \eqref{eq:y}. For each grid, 29,952 normal measurements for the first 312 days of one year are grouped as training data, and the remaining 54 days' normal measurements (namely 5,184 samples) are utilized for testing. The SFDIA measurements are grouped in a similar way, e.g., 179,712 for training and 31,104 for testing (each type of SFDIA contains 10,368 testing samples, with 5,184 SFDIA samples by modifying $Vm$ and 5,184 SFDIA samples by modifying $Va$). The advantage of this way of data partitioning is that the testing data is completely fresh to the trained model so that the detection cases can realistically simulate the real-world situation. 
	
	\subsection{Performance Metrics}
	Three commonly used metrics were applied to assess the SFDIA detection \cite{RN2097, RN2090, RN2110}, which are expressed by:
	\begin{equation} \nonumber
		\begin{cases}
			& Precision = \dfrac{N_{tp}}{N_{tp}+N_{fp}},\\
			&Recall = \dfrac{N_{tp}}{N_{tp}+N_{fn}},\\
			&F_1 =2\times \dfrac{Precision \times Recall}{Precision+Recall},
		\end{cases}
	\end{equation}
	where $N_{fp}$ indicates the number of false positive, $N_{tp}$ the number of true positive, $N_{fn}$ the number of false negative, and $N_{tn}$ the number of true negative, which are summarized in Table \ref{table:metric}. A normal measurement sample is defined as negative, while a sample attacked by the SFDIA is defined as positive.
	Hence, $N_{fn}+N_{tp}$ is the total number of real positive samples in the data set, and $N_{fp}+N_{tn}$ is the total number of real negative samples in the data set.
	\begin{table}[htb]
		\renewcommand{\arraystretch}{1.5}
		\caption{Definitions of performance metrics}
		\label{table:metric}
		\centering
		\begin{tabular}{c|c|c}
			\hline	
			\hline	
			\diagbox{Predicted result}{Genuine label }&Positive&Negative\\
			\hline
			Positive& $N_{tp}$& $N_{fn}$\\
			\hline
			Negative&$N_{fp}$&$N_{tn}$\\
			\hline
			\hline	
		\end{tabular}
		
	\end{table}
	
	%\subsection{Analysis and Validation of the proposed PowerFDNet}
	%measurement
	
	\subsection{Evaluation of SFDIA Detection}
	In the experiment, we compared the SFDIA detection accuracy of the proposed PowerFDNet with two state-of-the-art approaches, M-I \cite{RN2090} and M-II \cite{RN2110}. 
	As introduced in Section \ref{sec:SFDIAgeneration} and Section \ref{sec:trainingAndTesting}, there are 5,184 normal samples in the test stage. Each type of SFDIA contains 10,368 samples, where 5,184 samples are obtained by modifying the voltage magnitude of a target bus and the other 5,184 samples are obtained by modifying the voltage angle of a target bus. 
	%The results of SFDIA detection evaluated on Grid-HV and Grid-EHV are summarized in Table \ref{tab:type-I-Grid-HV}, Table \ref{tab:type-I-Grid-EHV}, Table \ref{tab:type-II-Grid-HV}, Table \ref{tab:type-II-Grid-EHV}, Table \ref{tab:type-III-Grid-HV}, and Table \ref{tab:type-III-Grid-EHV}. 
	Table \ref{tab:type-I-Grid-HV} and Table \ref{tab:type-I-Grid-EHV} compare the accuracy of Type-A SFDIA detection evaluated on Grid-HV and Grid-EHV, respectively. 
	Table \ref{tab:type-II-Grid-HV} and Table \ref{tab:type-II-Grid-EHV} compare the accuracy of Type-B SFDIA detection evaluated on Grid-HV and Grid-EHV, respectively.
	Table \ref{tab:type-III-Grid-HV} and Table \ref{tab:type-III-Grid-EHV} compare the accuracy of Type-C SFDIA detection evaluated on Grid-HV and Grid-EHV, respectively.
	The values in bold are the best results obtained for each accuracy. As shown in the tables, compared to the other two approaches, the proposed PowerFDNet has achieved significant improvements in the three performance metrics on the two benchmark power grids. 
	
	\subsubsection{Case A: Type-A SFDIA Detection}
	The Type-A SFDIA detection is to assess the detection accuracy of SFDIAs that are launched by attacking the target bus with a change in the bus active power injection in the range of $(50\%, 100\%]$.
	Table \ref{tab:type-I-Grid-HV} and Table \ref{tab:type-I-Grid-EHV} summarize the comparison of the detection accuracy of the Type-A SFDIAs evaluated on Grid-HV and Grid-EHV, respectively.
	\begin{table}[htbp]
		\setlength\tabcolsep{10pt}
		\renewcommand{\arraystretch}{1.5}
		\caption{Comparison of the detection accuracy of the Type-A SFDIAs evaluated on Grid-HV between the proposed PowerFDNet with the two state-of-the-art approaches.}
		\label{tab:type-I-Grid-HV}
		\centering
		\begin{tabular}{c||c|c|c} 
			\hline
			\hline
			\multirow{2}{*}{\textbf{Approaches}}
			&\multicolumn{3}{c}{\textbf{Grid-HV}}  
			\\ \cline{2-4} 
			&$Precision$ (\%) &$Recall$ (\%)&$F_1$ (\%)
			\\ \hline
			
			M-I \cite{RN2090}
			%			&96.147&92.670&94.377
			& 95.954 & 94.232 & 95.085
			\\ \hline%\Xhline{0.6pt}
			
			M-II \cite{RN2110} 
			&96.572 & 95.930 & 96.250
			\\ \hline%\Xhline{0.6pt}
			
			PowerFDNet
			& \textbf{99.557}&\textbf{99.778}&\textbf{99.668}
			\\ 	\hline
			\hline
		\end{tabular}
		%	\vspace{-0.1in}
	\end{table}
	
	\begin{table}[htbp]
		%	\vspace{-0.2in}
		\setlength\tabcolsep{10pt}
		\renewcommand{\arraystretch}{1.5}
		\caption{Comparison of the detection accuracy of the Type-A SFDIAs evaluated on Grid-EHV between the proposed PowerFDNet with the two state-of-the-art approaches.}
		%	\vspace{0.1in}
		\label{tab:type-I-Grid-EHV}
		\centering
		%	\vspace{-0.08in}
		\begin{tabular}{c||c|c|c} 
			\hline
			\hline
			\multirow{2}{*}{\textbf{Approaches}}
			&  \multicolumn{3}{c}{\textbf{Grid-EHV}}
			\\ \cline{2-4} 
			&$Precision$ (\%) &$Recall$ (\%)&$F_1$ (\%)
			\\ \hline
			
			M-I \cite{RN2090}
			&95.635 & 93.615 & 94.614
			\\ \hline%\Xhline{0.6pt}
			
			M-II \cite{RN2110} 
			&96.259 & 95.303 & 95.779
			\\ \hline%\Xhline{0.6pt}
			
			PowerFDNet
			& \textbf{99.422}&\textbf{99.508}&\textbf{99.465}
			\\ 	\hline
			\hline
		\end{tabular}
		%	\vspace{-0.1in}
	\end{table}
	
	As shown in Table \ref{tab:type-I-Grid-HV}, in the case of the Type-A SFDIA detection, it is clear that the proposed PowerFDNet achieves the best $F_1$ of $99.668$\%, the best recall of $99.778$\%, and the best precision of $99.557$\% on Grid-HV. 
	The precision achieved by the proposed method is about 3.756\% higher than M-I and 3.091\% higher than M-II, respectively.
	The recall achieved by the proposed method is about 5.885\% higher than M-I and 4.012\% higher than M-II, respectively.
	The $F_1$ score achieved by the proposed method is about 4.819\% higher than M-I and 3.551\% higher than M-II, respectively.
	Similar detection performance is also achieved on Grid-EHV, as summarized in Table \ref{tab:type-I-Grid-EHV}. 
	Our method obtains the best $F_1$ of approximately $99.465$\%, which is around 5.127\% higher than M-I and about 3.849\% higher than M-II, respectively. 
	The best precision obtained by our method is approximately $99.422$\%, which is about 3.960\% higher than M-I and 3.286\% higher than M-II, respectively.
	The best recall achieved by our method is approximately $99.508$\%, which is about 6.295\% higher than M-I and 4.413\% higher than M-II, respectively.
	That demonstrates that for the Type-A SFDIAs with large rates of power injection change at target buses, the PowerFDNet achieved the highest SFDIA detection accuracy in terms of the precision, recall, and $F_1$ score when compared to the two state-of-the-art approaches.

	\subsubsection{Case B: Type-B SFDIA Detection}
	The Type-B SFDIA detection is to assess the detection accuracy of SFDIAs that are launched by attacking the target bus through a medium modification in the range of $(25\%, 50\%]$ to the bus active power injection.
	Because there is less modification in the power injection in the Type-B SFDIAs, it is harder to detect the Type-B SFDIAs than to detect the Type-A SFDIAs. 
	Table \ref{tab:type-II-Grid-HV} and Table \ref{tab:type-II-Grid-EHV} compare the detection accuracy of the Type-B SFDIAs evaluated on Grid-HV and Grid-EHV, respectively. 
	\begin{table}[htbp]
		\setlength\tabcolsep{10pt}
		\renewcommand{\arraystretch}{1.5}
		\caption{Comparison of the detection accuracy of the Type-B SFDIAs evaluated on Grid-HV between the proposed PowerFDNet with the two state-of-the-art approaches.}
		%	\vspace{0.1in}
		\label{tab:type-II-Grid-HV}
		\centering
		%	\vspace{-0.08in}
		\begin{tabular}{c||c|c|c} 
			\hline
			\hline
			\multirow{2}{*}{\textbf{Approaches}}
			&\multicolumn{3}{c}{\textbf{Grid-HV}}  
			\\ \cline{2-4} 
			&$Precision$ (\%) &$Recall$ (\%)&$F_1$ (\%)
			\\ \hline
			
			M-I \cite{RN2090}
			&95.851 & 94.030 & 94.932
			\\ \hline%\Xhline{0.6pt}
			
			M-II \cite{RN2110} 
			&96.412 & 95.621 & 96.015
			\\ \hline%\Xhline{0.6pt}
			
			PowerFDNet
			& \textbf{99.461}&\textbf{99.576}&\textbf{99.518}
			\\ 	\hline
			\hline
		\end{tabular}
		%	\vspace{-0.1in}
	\end{table}
	
	\begin{table}[htbp]
		\setlength\tabcolsep{10pt}
		\renewcommand{\arraystretch}{1.5}
		\caption{Comparison of the detection accuracy of the Type-B SFDIAs evaluated on Grid-EHV between the proposed PowerFDNet with the two state-of-the-art approaches.}
		%	\vspace{0.1in}
		\label{tab:type-II-Grid-EHV}
		\centering
		%	\vspace{-0.08in}
		\begin{tabular}{c||c|c|c} 
			\hline
			\hline
			\multirow{2}{*}{\textbf{Approaches}}
			&  \multicolumn{3}{c}{\textbf{Grid-EHV}}
			\\ \cline{2-4} 
			&$Precision$ (\%) &$Recall$ (\%)&$F_1$ (\%)
			\\ \hline
			
			M-I \cite{RN2090}
			&95.541 & 93.200 & 94.356
			\\ \hline%\Xhline{0.6pt}
			
			M-II \cite{RN2110} 
			&96.170 & 94.927 & 95.544
			\\ \hline%\Xhline{0.6pt}
			
			PowerFDNet
			& \textbf{99.363}&\textbf{99.296}&\textbf{99.329}
			\\ 	\hline
			\hline
		\end{tabular}
		%	\vspace{-0.1in}
	\end{table}
	
	As shown in Table \ref{tab:type-II-Grid-HV}, in the case of Type-B SFDIA detection, it is clear to see that the proposed PowerFDNet obtains the best $F_1$ of $99.518$\%, the best precision of $99.461$\%, and the best recall of $99.576$\% on Grid-HV. 
	The precision achieved by our method is about 3.766\% higher than M-I and 3.162\% higher than M-II, respectively.
	The recall obtained by our method is about 5.898\% higher than M-I and 4.136\% higher than M-II, respectively.
	The $F_1$ score obtained by our method is approximately 4.831\% higher than M-I and 3.649\% higher than M-II, respectively.
	Similar detection performance is also achieved on Grid-EHV, as summarized in Table \ref{tab:type-II-Grid-EHV}. 
	Our method obtains the best $F_1$ of approximately $99.329$\%, which is around 5.271\% higher than M-I and about 3.962\% higher than M-II, respectively. 
	The best precision obtained by our method is approximately $99.363$\%, which is about 4.001\% higher than M-I and 3.321\% higher than M-II, respectively.
	The best recall achieved by our method is approximately $99.296$\%, which is about 6.540\% higher than M-I and 4.603\% higher than M-II, respectively.
	
	\subsubsection{Case C: Type-C SFDIA Detection}
	The Type-C SFDIA detection is to assess the detection accuracy of SFDIAs that are launched by attacking the target bus through a relatively small modification in the range of $(5\%, 25\%]$ to the bus active power injection.
	Compared with the other two SFDIAs, the Type-C SFDIA measurements have a smaller modification in the bus active power. Hence, it is more difficult to detect the Type-C SFDIAs. 
	Table \ref{tab:type-III-Grid-HV} and Table \ref{tab:type-III-Grid-EHV} summarize the detection accuracy of the Type-C SFDIAs evaluated on Grid-HV and Grid-EHV, respectively. 
	As clearly shown in Table \ref{tab:type-III-Grid-HV} and Table \ref{tab:type-III-Grid-EHV}, the detection accuracy (precision, recall, and $F_1$) of all the three methods is slightly lower than that evaluated on the Type-A and Type-B SFDIAs. 
	Compared to the other two methods, the proposed PowerFDNet achieved the highest detection accuracy in terms of $F_1$, recall, and precision on Grid-HV and Grid-EHV.
	
	\begin{table}[htbp]
		%	\vspace{-0.2in}
		\setlength\tabcolsep{10pt}
		\renewcommand{\arraystretch}{1.5}
		\caption{Comparison of the detection accuracy of the Type-C SFDIAs evaluated on Grid-HV between the proposed PowerFDNet with the two state-of-the-art approaches.}
		%	\vspace{0.1in}
		\label{tab:type-III-Grid-HV}
		\centering
		%	\vspace{-0.08in}
		\begin{tabular}{c||c|c|c} 
			\hline
			\hline
			\multirow{2}{*}{\textbf{Approaches}}
			&  \multicolumn{3}{c}{\textbf{Grid-HV}}
			\\ \cline{2-4} 
			&$Precision$ (\%) &$Recall$ (\%)&$F_1$ (\%)
			\\ \hline
			
			M-I \cite{RN2090}
			&95.748 & 93.837 & 94.783
			\\ \hline%\Xhline{0.6pt}
			
			M-II \cite{RN2110} 
			&96.311 & 95.428 & 95.867
			\\ \hline%\Xhline{0.6pt}
			
			PowerFDNet
			& \textbf{99.373}&\textbf{99.402}&\textbf{99.388}
			\\ 	\hline
			\hline
		\end{tabular}
		%	\vspace{-0.1in}
	\end{table}
	\begin{table}[htbp]
		%	\vspace{-0.2in}
		\setlength\tabcolsep{10pt}
		\renewcommand{\arraystretch}{1.5}
		\caption{Comparison of the detection accuracy of the Type-C SFDIAs evaluated on Grid-EHV between the proposed PowerFDNet with the two state-of-the-art approaches.}
		%	\vspace{0.1in}
		\label{tab:type-III-Grid-EHV}
		\centering
		%	\vspace{-0.08in}
		\begin{tabular}{c||c|c|c} 
			\hline
			\hline
			\multirow{2}{*}{\textbf{Approaches}}
			&  \multicolumn{3}{c}{\textbf{Grid-EHV}}
			\\ \cline{2-4} 
			&$Precision$ (\%) &$Recall$ (\%)&$F_1$ (\%)
			\\ \hline
			
			M-I \cite{RN2090}
			&95.423 & 92.892 & 94.140
			\\ \hline%\Xhline{0.6pt}
			
			M-II \cite{RN2110} 
			&96.092 & 94.637 & 95.359
			\\ \hline%\Xhline{0.6pt}
			
			PowerFDNet
			& \textbf{99.324}&\textbf{99.209}&\textbf{99.267}
			\\ 	\hline
			\hline
		\end{tabular}
		%	\vspace{-0.1in}
	\end{table}
	
	Compared to M-I evaluated on Grid-HV, our method improved by approximately 4.858\% in $F_1$ score, approximately 3.786\% in precision, and about5.931\% in the recall, respectively. 
	Compared with M-II evaluated on Grid-HV, our method obtained an improvement of approximately 3.672\%, 4.164\%, and 3.180\% in terms of $F_1$ score, recall, and precision, respectively.
	Compared with M-I evaluated on Grid-EHV, our method improved by approximately 5.446\% in $F_1$, about 6.801\% in the recall, and around 4.089\% in precision, respectively. 
	Compared with M-II evaluated on Grid-EHV, our method obtained an improvement of approximately 4.097\%, 4.831\%, and 3.363\% in terms of $F_1$, recall, and precision, respectively.
	That demonstrates that for the Type-C SFDIAs with small rates of power injection change at target buses, the PowerFDNet achieved the highest SFDIA detection accuracy ($F_1$, recall, and precision) compared with the two state-of-the-art approaches.
	
	\subsection{An IoT-oriented Prototype of the SFDIA detection}
\begin{figure}[htbp]
		\centering
		\includegraphics[width=0.5\linewidth]{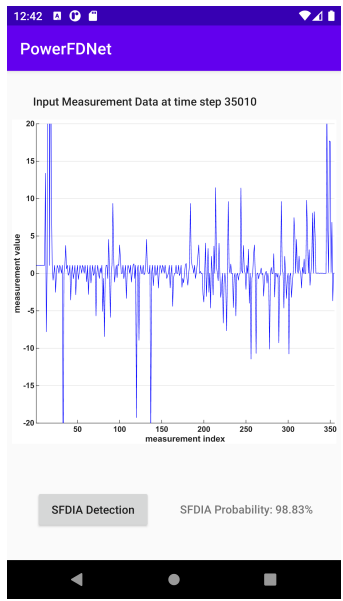}
		\caption{A screenshot of the IoT-oriented prototype implemented in the Android platform. It shows the detection result of an SFDIA measurement sample at time step 35010.}
		\label{fig:IoT_prototype}
		%	\vspace{-0.15in}
	\end{figure}

	A lightweight IoT-oriented SFDIA detection prototype was implemented in the Android-based mobile platform, as shown in Fig. \ref{fig:IoT_prototype}. The lightweight prototype is around of size 52 MB, with the optimized mobile model of size 8.5 MB. {\color{color} The optimized lightweight model is achieved by PyTorch, which provides such a utility to easily create serializable and optimizable models.\footnote{https://pytorch.org/mobile/android/\#quickstart-with-a-helloworld-example}} The testing time for one sample by the prototype is about 0.2 seconds in an android emulator of Pixel XL API 30. The popular deep learning framework PyTorch 1.10.0\footnote{https://pytorch.org/get-started/locally/} and PyTorch\_android\_lite:1.10.0\footnote{https://pytorch.org/mobile/android/} were used to implement this IoT-oriented prototype. 
	%		\begin{figure}[htbp]
	%		\centering
	%		\subfloat[][Detection on an SFDIA sample]
	%		{
	%			\centering
	%			\includegraphics[width=0.45\linewidth]{figures/iot_sfdia.eps}
	%			\label{fig:sfdia}
	%		}
	%		\hfil
	%		\subfloat[][Detection on a normal sample]
	%		{
	%			\centering
	%			\includegraphics[width=0.45\linewidth]{figures/iot_normal.eps}
	%			\label{fig:normal}
	%		}
	%		\caption[]{A screenshot of the IoT-oriented prototype implemented in the Android platform. \subref{fig:sfdia} shows the detection result of a SFDIA measurement sample at time step 600, and \subref{fig:normal} shows the detection result of the corresponding normal sample.}
	%		\label{fig:IoT_prototype}
	%	\end{figure}
	
	\section{Conclusion} \label{sec:conclusion}
	In this paper, we proposed a spatiotemporal deep learning-based PowerFDNet for successful SFDIA detection in AC-model power systems. 
	To model the spatiotemporal structure information between buses and lines, we designed two sub-architectures: the SA for the spatial structure learning and the TA for the temporal structure learning. 
	In the SA, we firstly model the bus measurements and the line measurements separately, so that the model can effectively represent these two types of measurements. 
	Then, a sub-network is designed to capture the spatial structure information between buses and lines and to preliminarily capture the patterns of SFDIA measurements. 
	Further, the TA based on the LSTM is designed to effectively learn the temporal structure information of the preliminary features obtained by the SA. 
	The proposed PowerFDNet is comprehensively evaluated on two realistic benchmark power grids. 
	The experimental results demonstrate that the PowerFDNet achieves significant improvement in terms of $F_1$, recall, and precision compared with the two state-of-the-art SFDIA detection approaches.
	In addition, an IoT-oriented lightweight prototype of size 52 MB is implemented and tested for mobile devices, which demonstrates the potential applications on mobile devices.
	
	\section*{Acknowledgment}
	
	This research was undertaken with the assistance of resources and services from the National Computational Infrastructure (NCI), which is supported by the Australian Government.
	This project was supported by ARC Discovery Grant with project ID DP190103660 and ARC Linkage Grant with project ID LP180100663.
	
	\ifCLASSOPTIONcaptionsoff
	\newpage
	\fi
	
	\bibliographystyle{IEEEtran}
	\bibliography{OJCS_2022}
	
	\vspace{-0.5in}
	\begin{IEEEbiography}
		[{\includegraphics[width=1in,height=1.25in,clip]{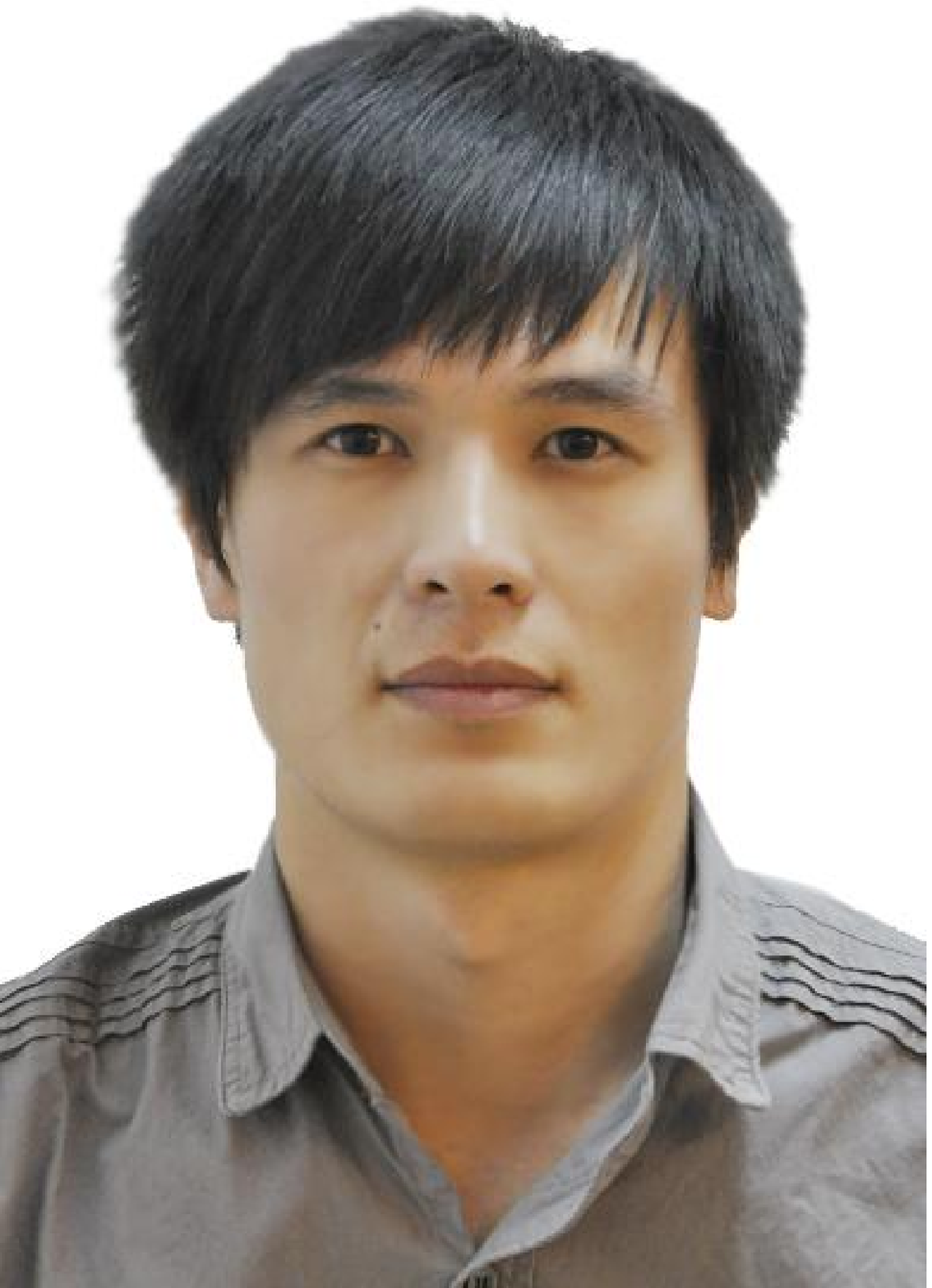}}]{Xuefei Yin}
		received the B.S. degree from Liaoning University, Liaoning, China; the M.E. degree from Tianjin University, Tianjin, China; and the Ph.D. degree from the University of New South Wales, Canberra, Australia. He is currently a Research Associate at the University of New South Wales in Canberra, Australia. His research interests include biometrics, pattern recognition, privacy-preserving, and intrusion detection. He has published articles in top journals including IEEE Transactions on Pattern Analysis and Machine Intelligence, IEEE Transactions on Information Forensics and Security, ACM Computing Surveys, and IEEE Transactions on Industrial Informatics.
	\end{IEEEbiography}
	
	\vspace{-0.5in}
	\begin{IEEEbiography}
		[{\includegraphics[width=1in,height=1.25in,clip]{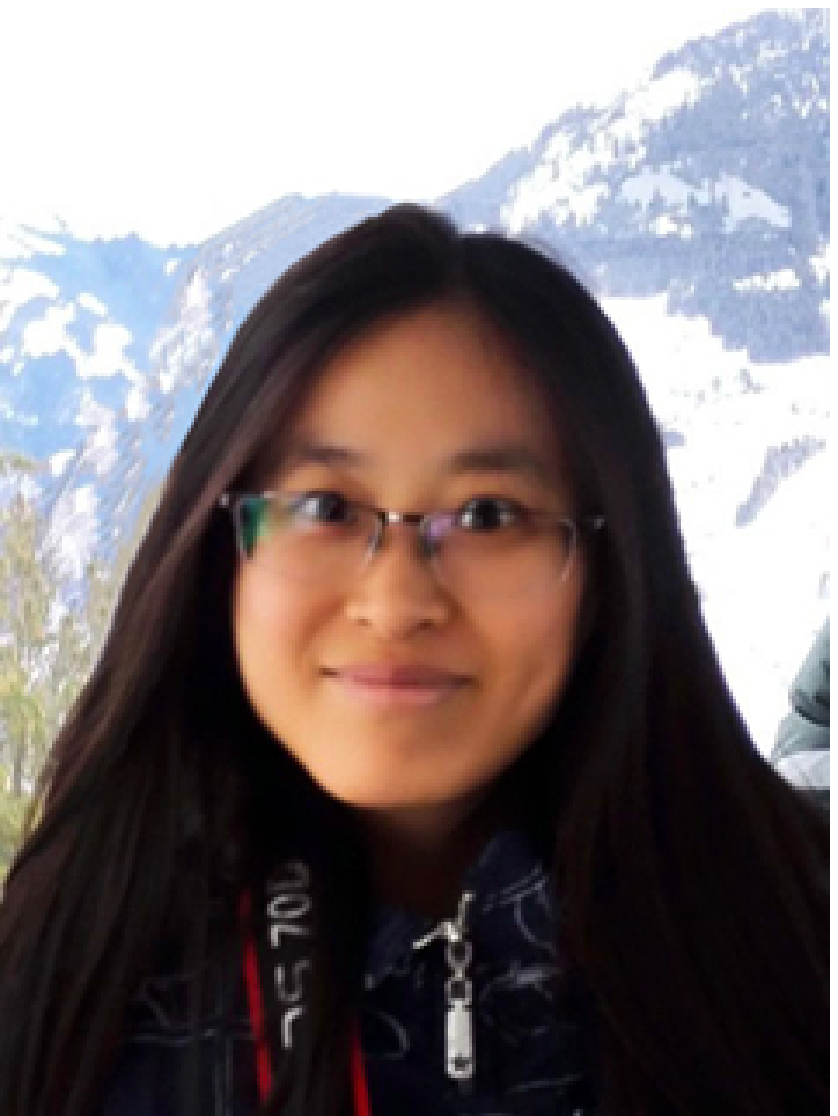}}]
		{Yanming Zhu}
		received the B.E. degree from Shandong Agricultural University, China; the M.E. degree from Tianjin University, China; and the Ph.D. degree from the University of New South Wales, Australia in 2019. She is currently a Research Fellow at the University of New South Wales, Sydney, Australia. Her research interests include deep learning, biometrics, and biomedical image analysis. She has published articles in top journals including IEEE Transactions on Pattern Analysis and Machine Intelligence, Pattern Recognition, IEEE Transactions on Information Forensics and Security, ACM Computing Surveys, and Bioinformatics.
	\end{IEEEbiography}
	
	\vspace{-0.5in}
	\begin{IEEEbiography}
		[{\includegraphics[width=1in,height=1.25in,clip]{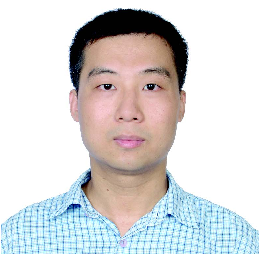}}]
		{Yi Xie}
		is currently an Associate Professor at the School of Data and Computer Science, Sun Yat-Sen University. He received the B.Sc., M.Sc. and Ph.D. degrees from Sun Yat-Sen University, Guangzhou, China. He was a visiting scholar at George Mason University and Deakin University during 2007 to 2008, and 2014 to 2015, respectively. He won the outstanding doctoral dissertation award of the Chinese Computer Federation (CCF) in 2009. His recent research interests include networking, cyber security and behavior modeling. Some of his works have been published in IEEE top journals, such as ToN, TPDS, TBD, TCSS and Sensors. He has received eight research grants and has served as a young Associate Editor for a Springer journal named Frontiers of Computer Science.
	\end{IEEEbiography}
	
	\vspace{-0.5in}
	\begin{IEEEbiography}
		[{\includegraphics[width=1in,height=1.25in,clip]{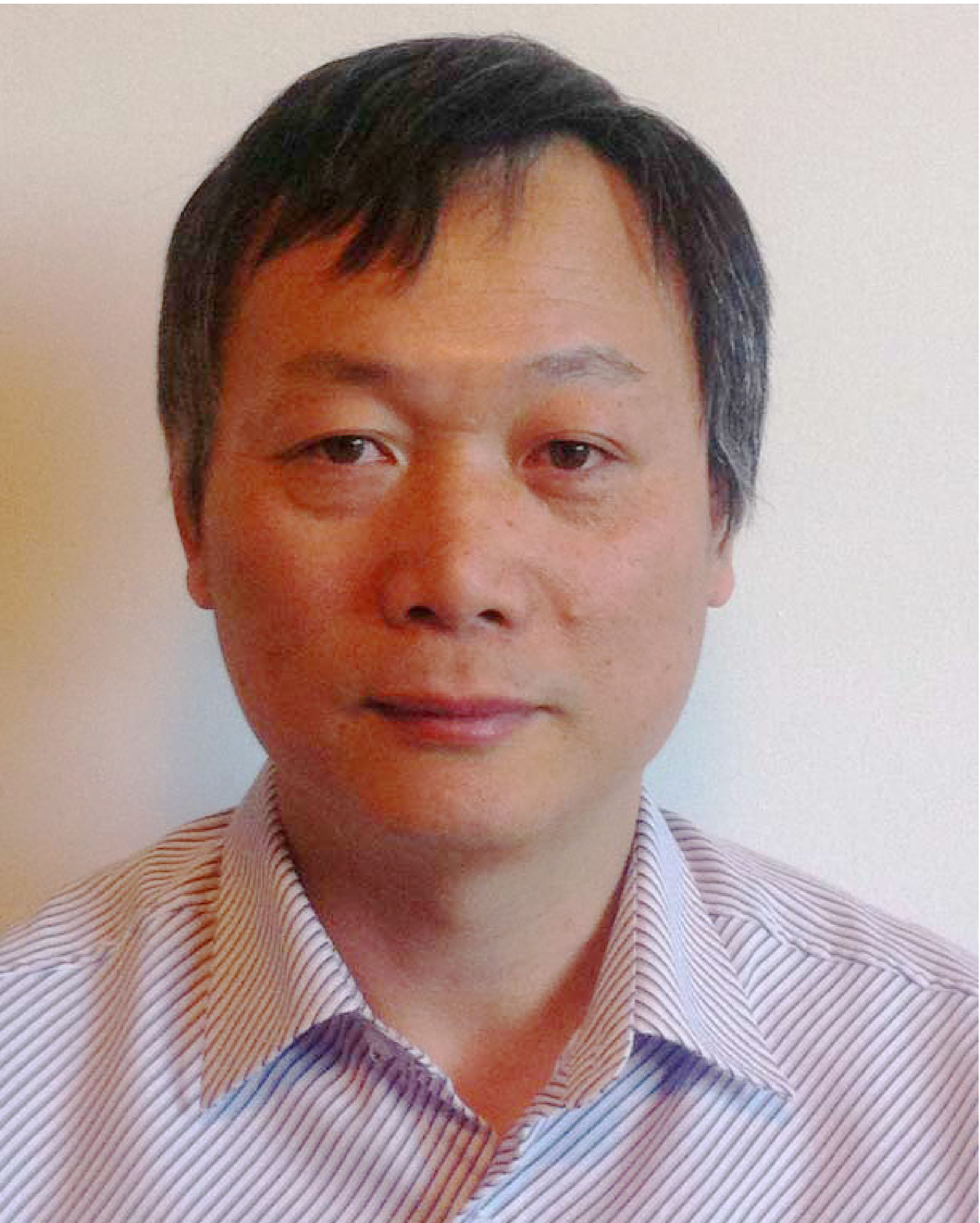}}]
		{Jiankun Hu}
		(Senior Member, IEEE) is currently a Full Professor with the School of Engineering and Information Technology, University of New South Wales, Canberra, Australia. He is an invited expert of the Australia Attorney-Generals Office assisting the draft of the Australia National Identity Management Policy. He has received nine Australian Research Council (ARC) Grants and has served at the Panel on Mathematics, Information and Computing Sciences, Australian Research Council ERA (The Excellence in Research for Australia) Evaluation Committee 2012. His research interest is in the field of cybersecurity covering intrusion detection, sensor key management, and biometrics authentication. He has published many articles in top venues including IEEE Transactions on Pattern Analysis and Machine Intelligence, IEEE Transactions on Computers, IEEE Transactions on Parallel and Distributed Systems, IEEE Transactions on Information Forensics and Security, Pattern Recognition, and IEEE Transactions on Industrial Informatics. He has served on the editorial board of up to seven international journals, including serving as Senior Area Editor for IEEE Transactions on Information Forensics and Security. He received ten Australian Research Council (ARC) grants and has also served for the prestigious Panel of Mathematics, Information and Computing Sciences (MIC), ARC ERA (The Excellence in Research for Australia) Evaluation Committee.
	\end{IEEEbiography}
	
	
	%\vfill
	

\end{document}